\documentclass[%
aps,
prc,
superscriptaddress,
amsmath,amssymb,
]{revtex4-1}
\usepackage[utf8]{inputenc}
\usepackage{graphicx}
\usepackage{dcolumn}
\usepackage{bm}
\usepackage{hyperref}
\hypersetup{colorlinks=true,
            allcolors=black,
            bookmarksopen=false,
            bookmarksnumbered
           }

\begin{document}

\title{Parity-Violating Inelastic Electron-Proton Scattering at Low $Q^2$ \\ Above the Resonance Region}

\author{D.~Androi\'c}
\affiliation{University of Zagreb, Zagreb, HR 10002, Croatia }
\author{D.S.~Armstrong}
\affiliation{William \& Mary, Williamsburg, VA 23187, USA}
\author{A.~Asaturyan}
\affiliation{A.~I.~Alikhanyan National Science Laboratory (Yerevan Physics Institute), Yerevan 0036, Armenia}
\author{K.~Bartlett}
\affiliation{William \& Mary, Williamsburg, VA 23187, USA}
\author{R.S.~Beminiwattha}
\affiliation{Ohio University, Athens, OH 45701, USA}
\affiliation{Louisiana Tech University, Ruston, LA 71272, USA}
\author{J.~Benesch}
\affiliation{Thomas Jefferson National Accelerator Facility, Newport News, VA 23606, USA}
\author{F.~Benmokhtar}
\affiliation{Christopher Newport University, Newport News, VA 23606, USA}
\author{J.~Birchall}
\affiliation{University of Manitoba, Winnipeg, MB R3T2N2, Canada}
\author{R.D.~Carlini}
\affiliation{Thomas Jefferson National Accelerator Facility, Newport News, VA 23606, USA}
\affiliation{William \& Mary, Williamsburg, VA 23187, USA}
\author{J.C.~Cornejo}
\affiliation{William \& Mary, Williamsburg, VA 23187, USA}
\author{S.~Covrig Dusa}
\affiliation{Thomas Jefferson National Accelerator Facility, Newport News, VA 23606, USA}
\author{M.M.~Dalton}
\affiliation{University of Virginia,  Charlottesville, VA 22903, USA}
\affiliation{Thomas Jefferson National Accelerator Facility, Newport News, VA 23606, USA}
\author{C.A.~Davis}
\affiliation{TRIUMF, Vancouver, BC V6T2A3, Canada}
\author{W.~Deconinck}
\affiliation{William \& Mary, Williamsburg, VA 23187, USA}
\author{J.F.~Dowd}
\affiliation{William \& Mary, Williamsburg, VA 23187, USA}
\author{J.A.~Dunne}
\affiliation{Mississippi State University,  Mississippi State, MS 39762, USA}
\author{D.~Dutta}
\affiliation{Mississippi State University,  Mississippi State, MS 39762, USA}
\author{W.S.~Duvall}
\affiliation{Virginia Polytechnic Institute \& State University, Blacksburg, VA 24061, USA}
\author{M.~Elaasar}
\affiliation{Southern University at New Orleans, New Orleans, LA 70126, USA}
\author{W.R.~Falk}
\thanks{Deceased}
\affiliation{University of Manitoba, Winnipeg, MB R3T2N2, Canada}
\author{J.M.~Finn}
\thanks{Deceased}
\affiliation{William \& Mary, Williamsburg, VA 23187, USA}
\author{C.~Gal}
\affiliation{University of Virginia,  Charlottesville, VA 22903, USA}
\author{D.~Gaskell}
\affiliation{Thomas Jefferson National Accelerator Facility, Newport News, VA 23606, USA}
\author{M.T.W.~Gericke}
\affiliation{University of Manitoba, Winnipeg, MB R3T2N2, Canada}
\author{J.~Grames}
\affiliation{Thomas Jefferson National Accelerator Facility, Newport News, VA 23606, USA}
\author{V.M.~Gray}
\affiliation{William \& Mary, Williamsburg, VA 23187, USA}
\author{K.~Grimm}
\affiliation{Louisiana Tech University, Ruston, LA 71272, USA}
\affiliation{William \& Mary, Williamsburg, VA 23187, USA}
\author{F.~Guo}
\affiliation{Massachusetts Institute of Technology,  Cambridge, MA 02139, USA}
\author{J.R.~Hoskins}
\affiliation{William \& Mary, Williamsburg, VA 23187, USA}
\author{D.~Jones}
\affiliation{University of Virginia,  Charlottesville, VA 22903, USA}
\author{M.K.~Jones}
\affiliation{Thomas Jefferson National Accelerator Facility, Newport News, VA 23606, USA}
\author{R.T.~Jones}
\affiliation{University of Connecticut,  Storrs-Mansfield, CT 06269, USA}
\author{M.~Kargiantoulakis}
\affiliation{University of Virginia,  Charlottesville, VA 22903, USA}
\author{P.M.~King}
\affiliation{Ohio University, Athens, OH 45701, USA}
\author{E.~Korkmaz}
\affiliation{University of Northern British Columbia, Prince George, BC V2N4Z9, Canada}
\author{S.~Kowalski}
\affiliation{Massachusetts Institute of Technology,  Cambridge, MA 02139, USA}
\author{J.~Leacock}
\affiliation{Virginia Polytechnic Institute \& State University, Blacksburg, VA 24061, USA}
\author{J.P.~Leckey}
\affiliation{William \& Mary, Williamsburg, VA 23187, USA}
\author{A.R.~Lee}
\affiliation{Virginia Polytechnic Institute \& State University, Blacksburg, VA 24061, USA}
\author{J.H.~Lee}
\affiliation{William \& Mary, Williamsburg, VA 23187, USA}
\affiliation{Ohio University, Athens, OH 45701, USA}
\author{L.~Lee}
\affiliation{TRIUMF, Vancouver, BC V6T2A3, Canada}
\affiliation{University of Manitoba, Winnipeg, MB R3T2N2, Canada}
\author{S.~MacEwan}
\affiliation{University of Manitoba, Winnipeg, MB R3T2N2, Canada}
\author{D.~Mack}
\affiliation{Thomas Jefferson National Accelerator Facility, Newport News, VA 23606, USA}
\author{J.A.~Magee}
\affiliation{William \& Mary, Williamsburg, VA 23187, USA}
\author{J.~Mammei}
\affiliation{Virginia Polytechnic Institute \& State University, Blacksburg, VA 24061, USA}
\affiliation{University of Manitoba, Winnipeg, MB R3T2N2, Canada}
\author{J.W.~Martin}
\affiliation{University of Winnipeg, Winnipeg, MB R3B2E9, Canada}
\author{M.J.~McHugh}
\affiliation{George Washington University, Washington, DC 20052, USA}
\author{D.~Meekins}
\affiliation{Thomas Jefferson National Accelerator Facility, Newport News, VA 23606, USA}
\author{K.E.~Mesick}
\affiliation{George Washington University, Washington, DC 20052, USA}
\affiliation{Rutgers, the State University of New Jersey, Piscataway, NJ 08854, USA}
\author{R.~Michaels}
\affiliation{Thomas Jefferson National Accelerator Facility, Newport News, VA 23606, USA}
\author{A.~Micherdzinska}
\affiliation{George Washington University, Washington, DC 20052, USA}
\author{A.~Mkrtchyan}
\affiliation{A.~I.~Alikhanyan National Science Laboratory (Yerevan Physics Institute), Yerevan 0036, Armenia}
\author{H.~Mkrtchyan}
\affiliation{A.~I.~Alikhanyan National Science Laboratory (Yerevan Physics Institute), Yerevan 0036, Armenia}
\author{N.~Morgan}
\affiliation{Virginia Polytechnic Institute \& State University, Blacksburg, VA 24061, USA}
\author{A.~Narayan}
\affiliation{Mississippi State University,  Mississippi State, MS 39762, USA}
\author{L.Z.~Ndukum}
\affiliation{Mississippi State University,  Mississippi State, MS 39762, USA}
\author{V.~Nelyubin}
\affiliation{University of Virginia,  Charlottesville, VA 22903, USA}
\author{W.T.H van Oers}
\affiliation{TRIUMF, Vancouver, BC V6T2A3, Canada}
\affiliation{University of Manitoba, Winnipeg, MB R3T2N2, Canada}
\author{V.F.~Owen}
\affiliation{William \& Mary, Williamsburg, VA 23187, USA}
\author{S.A.~Page}
\affiliation{University of Manitoba, Winnipeg, MB R3T2N2, Canada}
\author{J.~Pan}
\affiliation{University of Manitoba, Winnipeg, MB R3T2N2, Canada}
\author{K.D.~Paschke}
\affiliation{University of Virginia,  Charlottesville, VA 22903, USA}
\author{S.K.~Phillips}
\affiliation{University of New Hampshire, Durham, NH 03824, USA}
\author{M.L.~Pitt}
\affiliation{Virginia Polytechnic Institute \& State University, Blacksburg, VA 24061, USA}
\author{R.W.~Radloff}
\affiliation{Ohio University, Athens, OH 45701, USA}
\author{W.D.~Ramsay}
\affiliation{TRIUMF, Vancouver, BC V6T2A3, Canada}
\affiliation{University of Manitoba, Winnipeg, MB R3T2N2, Canada}
\author{J.~Roche}
\affiliation{Ohio University, Athens, OH 45701, USA}
\author{B.~Sawatzky}
\affiliation{Thomas Jefferson National Accelerator Facility, Newport News, VA 23606, USA}
\author{T.~Seva}
\affiliation{University of Zagreb, Zagreb, HR 10002, Croatia }
\author{M.H.~Shabestari}
\affiliation{Mississippi State University,  Mississippi State, MS 39762, USA}
\author{R.~Silwal}
\affiliation{University of Virginia,  Charlottesville, VA 22903, USA}
\author{N.~Simicevic}
\affiliation{Louisiana Tech University, Ruston, LA 71272, USA}
\author{G.R.~Smith}
\affiliation{Thomas Jefferson National Accelerator Facility, Newport News, VA 23606, USA}
\author{P.~Solvignon}
\thanks{Deceased}
\affiliation{Thomas Jefferson National Accelerator Facility, Newport News, VA 23606, USA}
\author{D.T.~Spayde}
\affiliation{Hendrix College, Conway, AR 72032, USA}
\author{A.~Subedi}
\affiliation{Mississippi State University,  Mississippi State, MS 39762, USA}
\author{R.~Suleiman}
\affiliation{Thomas Jefferson National Accelerator Facility, Newport News, VA 23606, USA}
\author{V.~Tadevosyan}
\affiliation{A.~I.~Alikhanyan National Science Laboratory (Yerevan Physics Institute), Yerevan 0036, Armenia}
\author{B.~Waidyawansa}
\affiliation{Ohio University, Athens, OH 45701, USA}
\author{P.~Wang}
\affiliation{University of Manitoba, Winnipeg, MB R3T2N2, Canada}
\author{S.P.~Wells}
\affiliation{Louisiana Tech University, Ruston, LA 71272, USA}
\author{S.A.~Wood}
\affiliation{Thomas Jefferson National Accelerator Facility, Newport News, VA 23606, USA}
\author{S.~Yang}
\affiliation{William \& Mary, Williamsburg, VA 23187, USA}
\author{P.~Zang}
\affiliation{Syracuse University, Syracuse, NY 13244, USA}

\collaboration{Q$_{\text{weak}}$ Collaboration}


\begin{abstract}
  We report the measurement of the parity-violating asymmetry for the inelastic
 scattering of electrons from the proton, at $Q^2 = 0.082$~GeV$^2$ and
 $ W = 2.23$~GeV, above the resonance region. The result $A_{\rm
  Inel} = - 13.5 \pm 2.0 ({\rm stat}) \pm 3.9 ({\rm syst})$~ppm agrees
  with theoretical calculations, and helps to validate the
  modeling of the $\gamma Z$ interference structure functions
  $F_1^{\gamma Z}$ and $F_2^{\gamma Z}$ used in those calculations, which
  are also used for determination of the two-boson exchange $\gamma$-$Z$
  box diagram
  ($\Box_{\gamma Z}$) contribution
  to parity-violating elastic scattering measurements. 
  A positive parity-violating asymmetry for inclusive $\pi^-$ production
  was observed, as well as positive beam-normal single-spin asymmetry for scattered electrons and a negative beam-normal single-spin asymmetry for inclusive $\pi^-$ production.
\end{abstract}

\maketitle

\section{Motivation}\label{sec:motivation}

The importance of two-boson exchange processes (e.g., $\gamma \gamma, \gamma Z$) to precision
electromagnetic and electroweak physics has become increasingly apparent in recent years.
For example, it is now widely believed that two-photon exchange contributions
can explain much (perhaps all) of the striking difference in the proton form factor ratio
$G_E^p/G_M^p$ as extracted from cross sections using the Rosenbluth separation technique
and that obtained from recoil polarization measurements, e.g., \cite{Puckett:2017flj}
 (see \cite{Afanasev:2017gsk} for a recent review).
In electron scattering, two-photon box diagrams generate such observables as
beam-normal single-spin asymmetries \cite{Carlson:2007sp}
and target-normal single-spin asymmetries \cite{Koshchii:2018bog},
as well as differences between $e^-p$ and $e^+p$ scattering cross sections
\cite{Guichon:2003qm,Blunden:2003sp},
all of which have motivated a number of experiments \cite{Wells:2000rx,Maas:2004pd,Armstrong:2007vm,Androic:2011rh,Abrahamyan:2012cg,Rios:2017vsw,Esser:2018vdp,
Rachek:2014fam,Zhang:2015kna,Henderson:2016dea,Rimal:2016toz}.
Superallowed nuclear beta-decay measurements, which are critical
ingredients to tests of the unitarity of the CKM matrix, have $\gamma W$ and $WZ$ box diagrams
as their largest nucleus-independent radiative corrections
\cite{PhysRevLett.121.241804,PhysRevD.100.013001,PhysRevLett.123.042503}. 
Theoretical control of these diagrams is therefore highly desirable \cite{Seng:2019plg}.

A particular example of the relevance of two-boson exchange diagrams is the case of the
$\gamma Z$ box diagram in parity-violating electron scattering (PVES). In PVES,
longitudinally-polarized electrons scatter from an unpolarized target (a proton
in the present case), and electroweak interference generates an asymmetry
between the scattering cross section for right-handed ($\sigma_R$)  and left-handed
($\sigma_L$) electrons,
\begin{equation}
A_{\rm PV} = \frac{\sigma_R - \sigma_L}{\sigma_R + \sigma_L} \; .
\end{equation}

Elastic PVES on the proton has been used as a powerful low-energy test of the Standard Model
\cite{Qweak.Nature}, because at sufficiently small four-momentum transfer, and
for forward-angle scattering, this
asymmetry depends in a simple way on the proton's weak charge, $Q^p_W$, via
\begin{equation}
{A_{\rm PV}/A_0}  =
   Q_{W}^{p} + Q^2 B (Q^{2},\theta), \; \; \;  A_0 = \left[ \frac{- G_F Q^2}{4 \pi \alpha \sqrt{2}} \right] \; ,
\label{BTermEq}
\end{equation}
where 
$G_F$ is the Fermi constant, $\alpha$ is the fine structure constant, $-Q^2$ is the four-momentum transfer squared,
and $B(Q^2,\theta)$ encodes hadron structure effects. Within the framework of the Standard Model,
the proton's weak charge depends in turn on the weak mixing angle, $\theta_W$,
through the tree-level relation $Q_W^p= 1 -  4\sin^2 \theta_W$.

Including radiative corrections, we have ~\cite{Erler:2003yk}:
\begin{equation}
Q_W^p=\left( \rho + \Delta_e \right) \left( 1 -  4\sin^2 \theta_W(0) + \Delta^\prime_e \right) + \Box_{WW} + \Box_{ZZ}+\Box_{\gamma Z}(0) \; ,
\label{eq:s2tw}
\end{equation}
where $\theta_W(0)$ is the weak mixing angle at zero momentum
transfer. The electroweak radiative correction terms $\rho$, $\Delta_e$, and
$\Delta^\prime_e$ are under good theoretical control and have been
calculated to sufficient precision~\cite{Erler:2003yk} for interpretation of existing \cite{Qweak.Nature} and
planned \cite{Becker:2018gzk} measurements of the weak charge. Similarly, the box diagram terms
$\Box_{WW}$ and $\Box_{ZZ}$, which are amenable to evaluation via perturbative QCD, are known to adequate precision \cite{Musolf:1990ts}. Thus, the proton's weak charge $Q_W^p$ is well-predicted within the Standard Model and provides an excellent
low-energy avenue to search for new physics, motivating the recent Q$_{\rm weak}$ \cite{Qweak.Nature} and
future P2 \cite{Becker:2018gzk} experiments.

This satisfactory situation was upset when Gorchtein and Horowitz
~\cite{Gorchtein:2008px} revealed that the $\gamma Z$ box diagram
$\Box_{\gamma Z}$ (in particular, the term $\Box_{\gamma Z}^V$, the
piece which involves the axial electron current and the vector hadron
current) was strongly energy-dependent and therefore significantly larger
(at the
relevant beam energy scale) than had been claimed in earlier
estimates \cite{PhysRevD.68.016006}. They also showed that the uncertainty of
$\Box_{\gamma Z}^V$ was large enough to potentially add noticeably to the
projected uncertainty for the Q$_{\rm weak}$ measurement.

Following that initial work by Gorchtein and Horowitz, several different theoretical groups have
performed calculations of the $\Box_{\gamma Z}^V$ term. Gratifyingly, there is excellent agreement
on the size of $\Box_{\gamma Z}^V$ (as well as of $\Box_{\gamma Z}^A$) from all these calculations
\cite{Sibirtsev:2010zg,Gorchtein:2011mz,Rislow:2010vi,Rislow:2013vta,Blunden:2011rd,Hall:2013hta,Hall:2015loa,PhysRevD.100.053007},
although there is not yet consensus on the
size of the theoretical uncertainty on  $\Box_{\gamma Z}^V$.

The most important inputs to the calculations of the $\Box_{\gamma Z}^V$ contributions are the $\gamma Z$ interference structure functions
$F_1^{\gamma Z}$ and $F_2^{\gamma Z}$, which are functions of $Q^2$ and $W^2$ (or, equivalently, $Q^2$
and Bjorken $x$). Unfortunately, experimental
input for these structure functions is scarce, unlike their purely electromagnetic analogs $F_1^{\gamma}$ and
$F_2^{\gamma}$. While there have been extractions of $F_1^{\gamma Z}$ and $F_2^{\gamma Z}$ using neutral-current
deep-inelastic scattering (DIS) experiments at HERA \cite{Aaron:2012qi}, those data were at very high $Q^2$
($ > 60~{\rm GeV}^2$) and small Bjorken $x$, while the region of the dispersion integral that is important
for $\Box_{\gamma Z}^V$ calculation is high $x$ and low $Q^2$. The various $\Box_{\gamma Z}^V$ calculations
differ primarily in how the $F_1^{\gamma Z}$ and $F_2^{\gamma Z}$ were modelled in this kinematic
regime, and in the uncertainties ascribed to this modeling. 

This modeling of the $\gamma Z$ interference structure functions can
be tested by comparing to parity-violating electron scattering
data. However, there are only two previous PVES experiments that can
be used to constrain or test models of $F_1^{\gamma Z}$ and
$F_2^{\gamma Z}$. These are the measurement of parity-violating
inelastic scattering near the $\Delta(1232)$ resonance by the G0
collaboration, who extracted the parity-violating asymmetry from the
proton and deuteron at $Q^2 = 0.34~{\rm GeV}^2$ and $W = 1.18$~GeV
\cite{Androic:2012doa}, and the JLab Hall A E08-011 (PVDIS)
collaboration, which measured asymmetries from electron-deuteron
scattering over several values of $W$ between 1.2 to 1.98 GeV and
$Q^2$ between 0.95 and 1.47 GeV$^2$ \cite{Wang:2013kkc}. Constraints
based on the results from these two experiments were applied by the Adelaide-Jefferson Lab-Manitoba (AJM)
theoretical group \cite{Hall:2013hta}
and were important in reducing their uncertainty in $\Box_{\gamma Z}^V$.
The AJM group subsequently adopted quark-hadron duality parton distribution
function (PDF) fits in order to apply additional constraints on the interference
structure functions \cite{Hall:2015loa}.

Additional experimental input to test these models of the interference structure
functions, and thus test the calculation of the $\Box_{\gamma Z}^V$ diagram, would clearly be valuable;
this motivated the present measurement. During a special running period
of the Q$_{\rm weak}$ experiment~\cite{Qweak.Nature}, the beam energy was raised in order to
measure the parity-violating asymmetry from the proton in an inelastic
region of interest for the $\Box_{\gamma Z}^V$ calculations ($Q^2 = 0.082$~GeV$^2$ and $ W = 2.23$~GeV), and for which
asymmetry predictions were available using the structure function models from
two of the theoretical groups (AJM ~\cite{Hall:2013hta} and Gorchtein {\em et al.} \cite{Gorchtein:2011mz}). For this measurement the spectrometer accepted
electrons and pions with a scattered momentum between 900 and 1300 MeV and
an angle between $5.8^{\circ}$ and
$11.6^{\circ}$. For the electrons this corresponds to a range of $Q^2$ from
0.04 to 0.15~GeV$^2$ and $0.01 < x < 0.035$.

In the remainder of this paper, we describe this measurement and the data analysis, and compare
the extracted asymmetry to the model predictions.

\section{Experiment}\label{sec:experiment}

The measurement was conducted using the Q$_{\rm weak}$ apparatus, which was located in
Hall C at the Thomas Jefferson National Accelerator Facility.
This apparatus was optimized for the
measurement of the parity-violating asymmetry in the elastic
scattering of longitudinally-polarized electrons at a beam energy of 1.16 GeV from the protons in a liquid hydrogen target (the Q$_{\rm weak}$
weak-charge measurement~\cite{Qweak.Nature}). 
A detailed
discussion of the apparatus is available
elsewhere~\cite{ALLISON2015105}; here we provide an overview,
followed by a presentation of those aspects that were modified for the
present measurement.

The Q$_{\rm weak}$ apparatus was designed to detect electrons scattered from
the target with scattering polar angles between $5.8^{\circ}$ and
$11.6^{\circ}$ and 49\% of $2\pi$ in azimuth.  The angular
acceptance was defined by a sequence of three precision
lead-alloy collimators, each with eight symmetric apertures in
$\theta$ and $\phi$.  The azimuthal acceptances were matched to the
eight open sectors (``octants'') of a toroidal spectrometer
magnet. Momentum-selected scattered electrons were detected by one of
eight identical fused-silica Cherenkov detectors, arranged in an
azimuthally-symmetric array, one detector per magnet octant. These
detectors, called the main detectors (MD, numbered MD1 to MD8) were
rectangular bars 2 m in length, 18 cm in width and 1.25 cm in
thickness. Cherenkov light was read out from each MD using a pair of
photomultiplier tubes (PMTs), one located at each end of a given
detector. Just upstream of each MD was a 2-cm thick Pb
pre-radiator, which served to amplify the signal from incident
electrons through generation of an electromagnetic shower and to
suppress low-energy backgrounds.  The main detectors were enclosed in
a concrete shielding hut, with 122 cm thick walls. The upstream face
of the hut was constructed of 80 cm thick high-density (2700
kg/m$^{3}$) concrete, loaded with barite (BaSO$_4$).

In the standard mode of data-taking used for the asymmetry
measurements, referred to here as ``Integrating Mode'',
the current produced by each main detector PMT was
converted to a voltage and integrated during $\approx 1$~ms long periods,
known as ``helicity windows'', during which the electron beam helicity was held
constant (see below). The integrated PMT signal for each helicity
window was then digitized and recorded.

An alternate mode, ``Event Mode'', was used in which the beam current was reduced
by six orders of magnitude (to $\approx 0.1$~nA), and PMT signals caused by individual
scattering events were read out individually and digitized.
This enabled pulse-height and timing analysis of
individual scattered events, which was useful for determining
background fractions.  During Event-Mode data-taking, a set of drift
chambers, known as the tracking system, was inserted upstream and
downstream of the magnet to track individual charged particles during
dedicated periods of low-current running.  This system was used for
calibration purposes, background studies, and for confirmation of the
kinematics and acceptance of the detected electrons.

The polarized electron-beam’s helicity was selected 960 times per
second, allowing the beam to be produced in a sequence of ``helicity
quartets'', either $(+--+)$ or $(-++-)$, with the pattern chosen
pseudorandomly at 240 Hz.  The quartet pattern served to cancel
effects due to slow linear drifts in beam properties or detector
gains, while the rapid helicity reversal suppressed noise
contributions due to fluctuations in either the target density or beam
properties. An additional ``slow'' reversal of the helicity was done
every 4 hours by insertion or removal of a half-wave plate
in the path of the circularly-polarized laser beam used to generate
the polarized electron beam.

Continuous measurements of the incident electron-beam current were
made using three independent radio-frequency resonant-cavity beam
current monitors (BCMs). The beam trajectory was measured using five
beam-position monitors (BPMs) located upstream of the target. 
Energy changes were measured using another BPM at a dispersive
location in the beam line.

The primary target was a high-power cryogenic liquid-hydrogen
target. The hydrogen was maintained at 20 K and was contained in an
aluminum-walled target cell, 34.4 cm in length, with thin Al-alloy entrance
and exit windows (respectively, 0.11 mm and 0.13 mm thick). Several
additional solid targets were available, in particular a 3.7 mm thick
Al target, made of the same alloy as the hydrogen-target entrance
and exit windows; the thickness of the Al target was chosen to match
the radiation length of the hydrogen target.

For the present measurement, the apparatus and experimental conditions
were modified in two main ways, compared to the weak-charge measurement:
the beam energy was increased, and one main detector was modified to
have an increased sensitivity to pions. We discuss both of these
changes in the following paragraphs.

The beam energy was raised to 3.35~GeV
in order to access the inelastic scattering kinematics of interest.
The incident beam current was maintained between 160 and 180 $\mu$A.
Due to beam-delivery requirements for an experiment running
concurrently in another experimental hall, it was not possible to
deliver an electron beam which was polarized fully in the longitudinal
direction. Instead, the electron spin-angle during the main data-taking,
which we refer to as the ``mixed'' data set, was at $\theta_P^{\rm mix} = -19.7^{\circ}\pm 1.9^{\circ}$,
where a positive angle corresponds to an angle measured from the beam
axis, 
 rotated towards beam right in the horizontal plane.
This
corresponded to an electron spin with a 94.1\% longitudinal component
and a 33.7\% (horizontal) transverse component.  Beam of the same
energy, but polarized almost entirely in the horizontal transverse
orientation, with a polarization angle of $\theta_P^{\rm trans} =
92.2^{\circ} \pm 1.9^{\circ}$, was available for part of the
data-taking, which we refer to as the ``transverse'' data set.

The average magnitude of the polarization of the electron beam was
$P = 0.870 \pm 0.006$ for both mixed and transverse data, as
measured by the Compton and  M\o ller
polarimeters~\cite{Narayan:2015aua,Magee:2016xqx} in Hall C.

Due to the higher beam energy, a significant background was present in
the main detectors caused by negative pions produced in the target
with similar momenta to the inelastically scattered electrons of
interest. Positively charged pions were swept out of the acceptance by
the spectrometer magnetic field. With the high-rate integrating mode
of detector readout, it was not possible to separate the contributions
of individual electrons from individual pions to the asymmetry
measurement. In order to measure and correct for this pion background,
one of the main detectors (in octant 7) was modified so as to have an
enhanced sensitivity to pions. The modification was the addition of a
10.2~cm thick Pb absorber ($ \approx 18$ radiation lengths), placed just
upstream of the detector. This significantly attenuated the signal in
this detector from scattered electrons, without affecting the signal from
the majority of the $\approx 1$~GeV pions. Thus the asymmetry in MD7
was dominated by that from incident pions, with a different mixture of
electron and pion signals than in the other 7 main detectors.
Under the assumption that the pion and electron fluxes were
azimuthally symmetric, this difference allowed an unfolding of the
separate electron and pion asymmetries. We refer to the detectors
other than MD7 as ``unblocked'' detectors.

The mixed data set on the hydrogen target included $9.4\times
10^7$ helicity quartets (108 hours of data-taking), and the transverse
data set included $3.7\times 10^6$ helicity quartets (4.3 hours of
data-taking). The rate of charged particles incident on each unblocked
detector was approximately 9 MHz, of which approximately 27\% were
pions and 73\% electrons. As a typical electron produced significantly
more light in an unblocked detector than did a pion, the fraction of the
integrated detector signal due to pions (see Sect.~\ref{sec:dilutions}) was 9.6\%, with the remainder being mainly due to electrons. In the blocked 
detector, MD7, 39.7\% of the integrated detector signal was due to
pions, 9.3\% due to electrons, and 51\% due to neutral particles
(see Sect.~\ref{sec:neutral}).

In order to measure the asymmetry caused by electrons scattering from
the target entrance and exit windows, ``mixed'' data were also taken
on the Al alloy target; for these runs the beam current was reduced to
60 $\mu$A. This aluminum data set included $5.2\times 10^6$ helicity
quartets (6.0 hours of data-taking).

\section{Data Analysis}\label{sec:analysis}

The extraction of the parity-violating asymmetry for the inelastic
scattering events took place in several steps. Event-Mode data were used to
determine the fraction of the experimental yield that arose due to
pions and neutral particles (Sec.~\ref{sec:dilutions}). The
Integrating-Mode data were used to form asymmetries for each detector, and these
were then corrected for several classes of false asymmetry (Sec.~\ref{sec:asymmetry}). The resulting asymmetries and yield fractions for all eight detectors
were analyzed in a combined fit in order to separately extract
the transverse and longitudinal asymmetries for both electrons and
pions (Sec.~\ref{sec:pves}). Finally, the longitudinal electron asymmetry
$A_e^{L}$ obtained from this fit was corrected for the effect of various backgrounds
(Sec.~\ref{sec:pvis}) in order to obtain the inelastic parity-violating
asymmetry $A_{\rm Inel}$.

Each of these steps in the analysis is described in the following sections. 
Further details can be found in the PhD thesis of one of us~\cite{Dissertation:Jim}.

\subsection{Yield-Fraction Determination}\label{sec:dilutions}

Event-Mode data were used to determine the fraction of the yield
in the detectors that was caused by pions and by neutral particles
(as opposed to the desired electrons). The
existence of the Pb wall in front of the ``blocked'' detector (MD7) meant
that these determinations were done differently for that detector compared
to the seven other unblocked detectors, as discussed below.

\subsubsection{Pion yield-fraction in unblocked detectors}\label{sec:pionyield}

\begin{figure}[htb]
\begin{center}
\includegraphics[width=3in]{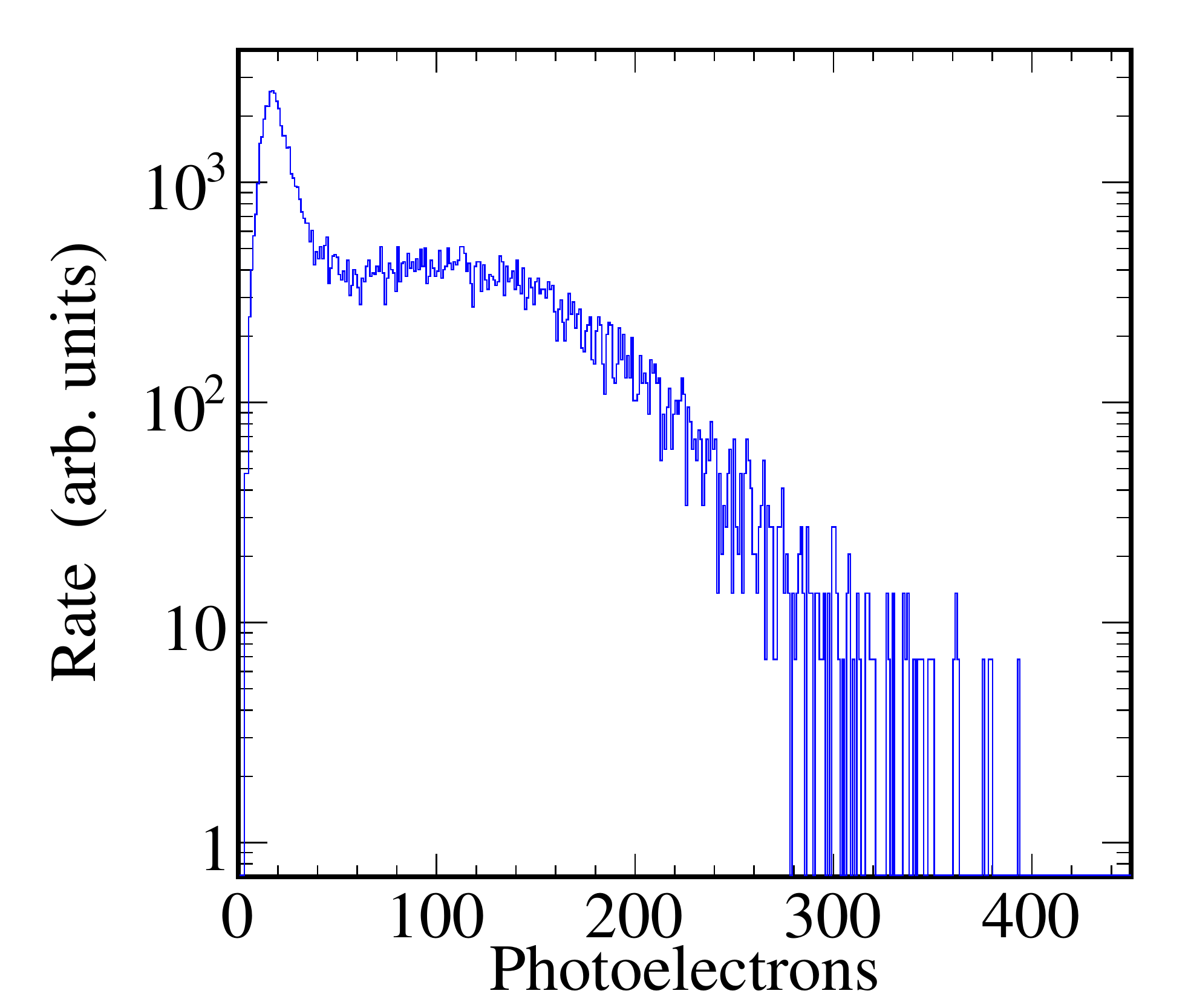}
\caption{Typical photoelectron spectrum, after subtraction of the electronic pedestal, for an unblocked main-detector (MD4). Two distinct peaks can be seen. The narrow peak on the left, centered near 20 photoelectrons, is due to pions.
The broader peak on the right, centered near 100 photoelectrons, is due to electrons. } \label{fig:simple_adc}
\end{center}
\end{figure}

Event-Mode data were used to determine $f_{\pi}$, the fraction of the
pulse-height weighted signal seen in the detectors that was due to pions
(or muons from their
decay; these were treated together, and henceforth ``pions'' will refer to both).
We note that
negative pions can only arise in electron scattering on hydrogen
through multi-pion production processes. There is a paucity of
multi-pion production cross-section data available in the relevant
kinematic range for this experiment, and so we relied on measured
pulse-height distributions in the main detectors to determine
$f_{\pi}$. Figure~\ref{fig:simple_adc} shows a typical pulse-height
spectrum in photoelectrons, for an unblocked main-detector, obtained during Event-Mode data-taking. The spectrum is a sum of the calibrated signals of the two PMTs that read out the detector.
The broad peak at larger pulse-height due to the showering
electrons is clearly seen, along with the peak at lower pulse-height due
to the (minimum-ionizing) pions.

\begin{figure}[htb]
\begin{center}
\includegraphics[width=3in]{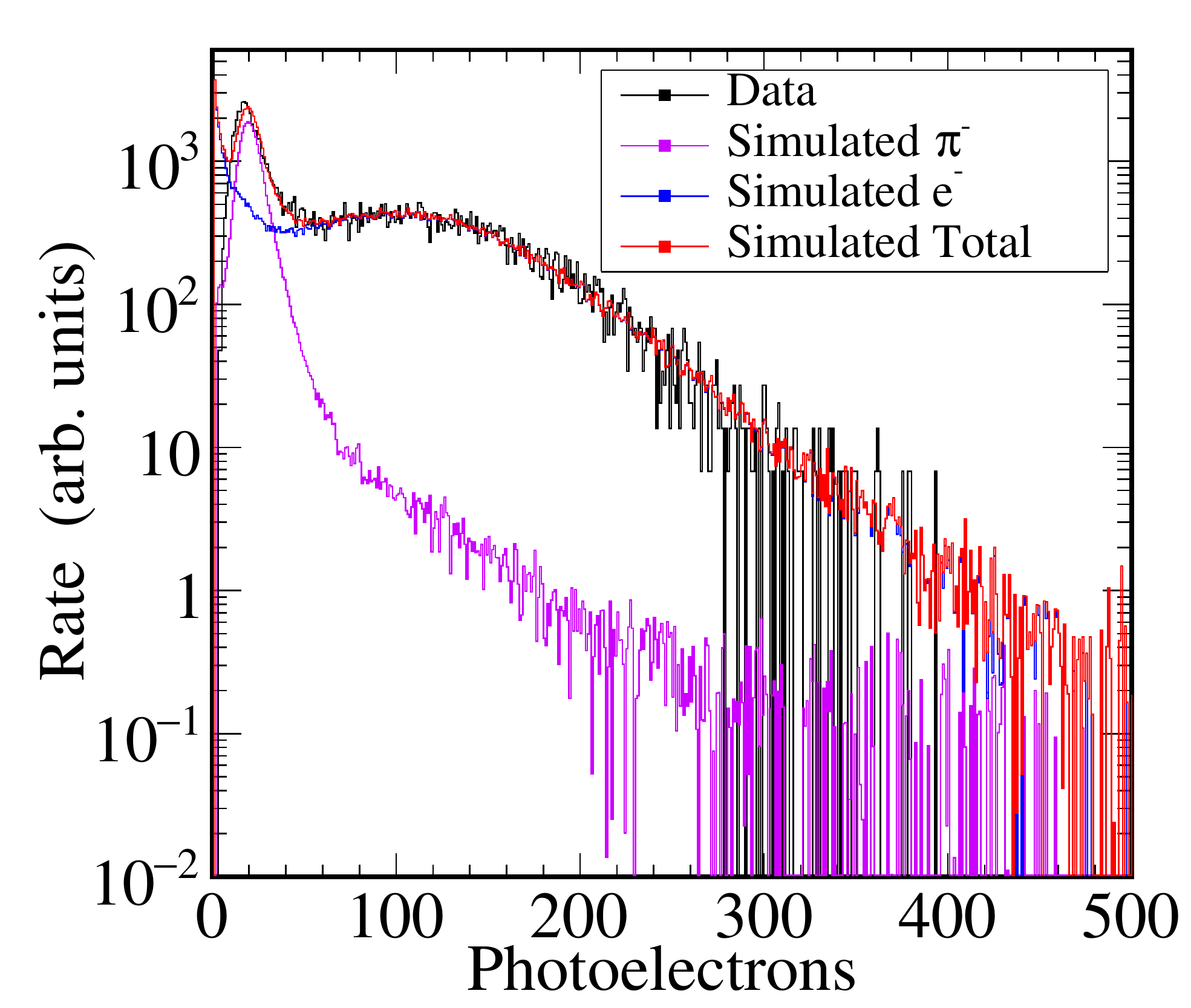}
\caption{Typical fit to the photoelectron spectrum from an unblocked main-detector (MD4). The black histogram is the data, and the simulated spectra from
  pions (magenta), electrons (blue) and the sum of the electrons and pions
  (red) are superimposed.} \label{fig:pe_fit}
\end{center}
\end{figure}

In order to fit these spectra to determine the fractional signal from
pions and electrons, separate Geant4~\cite{Agostinelli:2002hh} simulations were generated for
electrons and pions, with appropriate momenta, incident on a detector,
with the generation and tracking of optical photons enabled in the simulation.  These simulations
provided pulse-height distributions for the two particle types. The
experimental pulse-height spectra for each of the unblocked main-detectors were fit to a linear combination of the simulated pion
and electron spectra, with the relative fractions of the two particle
types as a free parameter. The fit also included a scaling factor
between photoelectrons (simulation) and electronic channels (data), one
factor for each detector, to account for detector-to-detector
variations in PMT gain. A typical fitted spectrum is shown in Fig.~\ref{fig:pe_fit}.

The pion yield-fraction was calculated as
  \begin{equation}
    f_{\pi}^i=\frac{Y_{\pi}^{\rm sim}}{Y_{\pi}^{\rm sim}+Y_e^{\rm sim}} \; ,
  \end{equation}
  where $i$ is the detector number, $Y_{\pi}^{\rm sim}$ is the total simulated
  light-yield from pions,
and $Y_e^{\rm sim}$ is the total simulated light-yield from electrons.

\begin{figure}[htb]
\begin{center}
\includegraphics[width=3in]{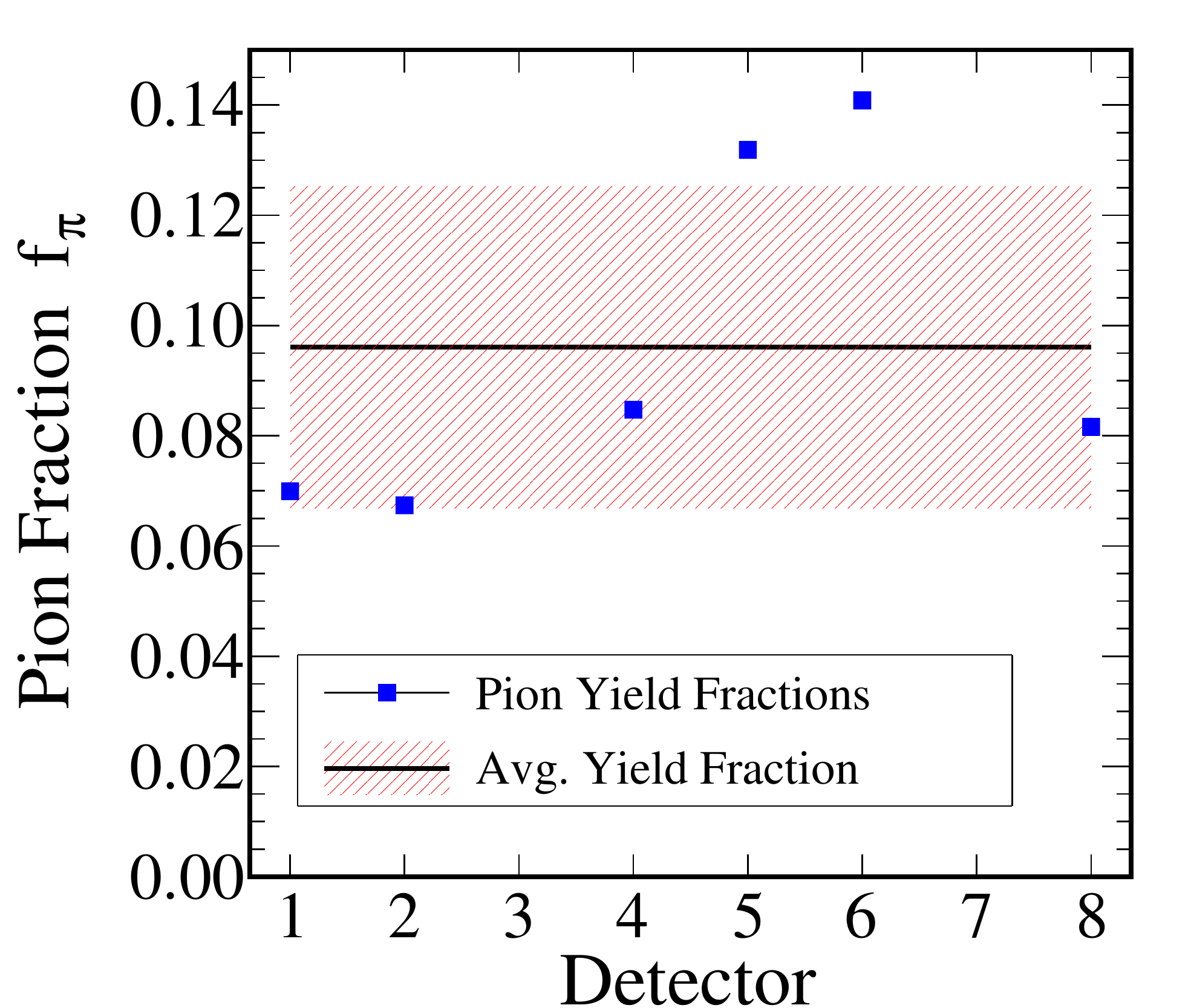}
\caption{The fraction of the detector yield due to pions, $f_{\pi}^i$, is shown for each
  main detector, except MD3 and MD7 (see text).
  The statistical uncertainties from the fitting routine are
  smaller than the plotting symbols. Also shown is the average
  value (black line) and the RMS (hatched area), which we adopt as the uncertainty on the average.} \label{fig:pion_fraction}
\end{center}
\end{figure}

In Figure~\ref{fig:pion_fraction}, the fitted and pulse-height weighted pion-fractions for six of the
eight main detectors are shown. The blocked detector
MD7 was not included in this analysis, because it required a different approach
to determine its pion yield-fraction (see next subsection). Main detector 3 was
also not included, due to a noisy PMT that
distorted the pulse-height spectrum in Event-Mode data. The Integrating-Mode
asymmetry data were taken at a rate that was six orders of magnitude higher, 
so that this noise was negligible compared to the signal in that case.
Thus, MD3 was
included in the asymmetry analysis (Sec.~\ref{sec:analysis}).
  
There was a significantly larger detector-to-detector variation in the pion fraction values
than one would expect from statistical uncertainty alone, presumably
due to an unaccounted-for systematic effect (we do not know of a physics
reason for such a variation).
The root-mean-squared deviation
(RMS) of the six $f_{\pi}^i$ was used as a conservative uncertainty on
the average pion yield-fraction to account for this systematic effect,
giving a detector-averaged value of $f_\pi^{\rm avg} = 0.096 \pm 0.029$ for
the unblocked detectors. 

\subsubsection{Pion yield-fraction in MD7}\label{sec:pionmd7}

\begin{figure}[htb]
\begin{center}
\includegraphics[width=3in]{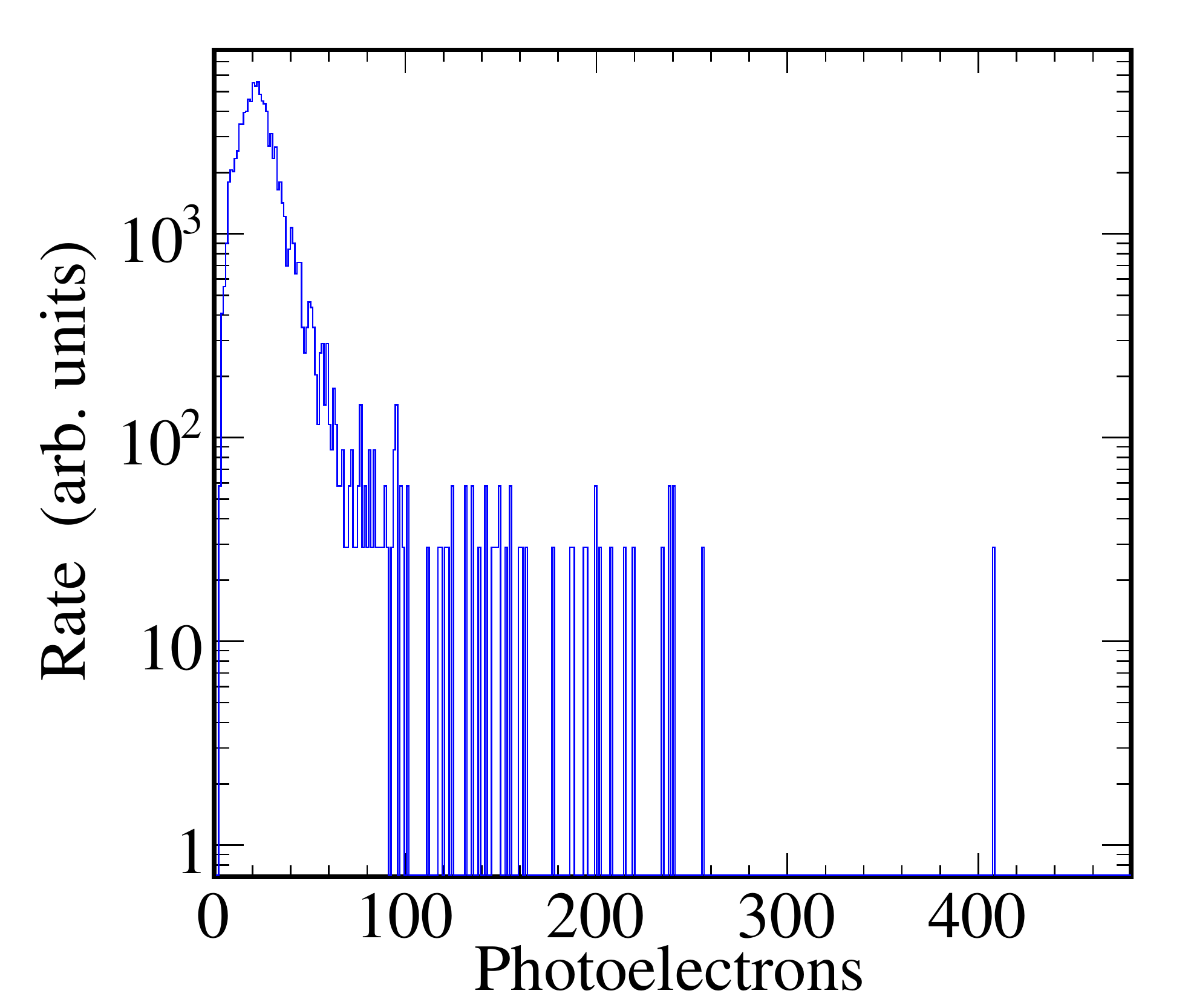}
\caption{Photoelectron spectrum from the blocked main detector (MD7).
  Only the peak due to pions (~$\approx 20$~photoelectrons) is visible;
  {\em c.f.}~Fig.~\ref{fig:simple_adc}.} \label{fig:simple_adc_md7}
\end{center}
\end{figure}

The method described above could not be used to determine the pion yield-fraction in MD7, the blocked detector, as the fraction of electrons
surviving the Pb wall was so small (see Fig.~\ref{fig:simple_adc_md7}) that the electron
peak could not be reliably  distinguished from the tail of the pion
peak in the pulse-height spectrum. Instead, the assumption was made that both the electrons and the pions were produced with approximate
azimuthal symmetry, so that the same pion/electron flux ratio was incident on all 8 detectors.
Additional Geant4 simulations were generated for
electrons and pions, of appropriate momenta, incident on the Pb wall
in MD7, to determine the signal-attenuation factors in the Pb for the
electrons and pions, respectively. Applying these attenuation factors
to the incident flux-ratio extracted from the unblocked detectors yielded
a pion light-yield fraction for MD7 of $f_{\pi}^7  = 0.81 \pm  0.05$.

\subsubsection{Neutral yield-fraction}\label{sec:neutral}

The Event-Mode data were also used to determine the fraction of the
main-detector signal which arose due to neutral particles (primarily
low-energy gamma rays, but also neutrons). During this data-taking,
incident charged-particles could be vetoed using the plastic
``trigger'' scintillators from the tracking system (for a detailed
description of the tracking system, see
Ref.~\cite{ALLISON2015105}). These scintillators (218 cm long, 30 cm wide and
1.0 cm thick), when moved into measurement position, were located just
upstream of the main detectors, and covered the entire
acceptance of the main detectors for particles from the target passing
through the spectrometer. For the measurement of the neutral
yield-fraction, the data acquisition was triggered by the main-detector
signal, and charged events were rejected in offline data analysis by
placing cuts on the time and pulse-height spectra from the
scintillators. The fraction of the yield in the unblocked main-detectors
that was due to neutral particles was measured to be
$(7.8\pm 0.4)$\%.  This was considerably larger than the $<0.3$\% fraction
observed in the weak-charge measurement~\cite{Qweak.Nature}.  This
large neutral-fraction was due to several effects: (i) the
contribution of ``punch-through'' events, in which $\approx 3$ GeV
elastically-scattered electrons showered in the detector-hut shield
wall, (ii) the much lower rate of inelastic electrons at the present
kinematics compared to the dominant elastic-electrons of the weak-charge
measurement, and (iii) the ``glow" of low-energy gamma rays from
the interaction of very-forward scattered electrons with the
beam-pipe downstream of the target.

The effect of the punch-through events on the asymmetry measurement
was corrected for separately (see Sec.~\ref{sec:pt}), so here Geant4
simulation was used to estimate the contribution of punch-through events to the
neutral yield-fraction. Subtracting this gives a non-punch-through
neutral fraction of $f_{\rm NB}^{\rm Un} = (6.3 \pm 0.6)$\%.
As this background is understood to arise mainly due to small-angle
events, {\em i.e.}, events at very low $Q^2$, it carries essentially no physics
asymmetry (parity-violating asymmetries generically scale with $Q^2$,
and parity-conserving asymmetries from two-photon exchange generically
scale with $Q$).
We therefore treat this as a pure ``dilution'' to the asymmetry
measurement. This background does carry a false asymmetry, $A_{\rm BB}$, the
correction for which is described in Sec.~\ref{sec:asymmetry}.

In the case of MD7, however, the presence of the Pb wall upstream of
the detector prevented the tracking system from being moved into place
to veto charged particles.  Therefore, a direct measurement of the
neutral yield-fraction for this detector, $f_{\rm NB}^7$, was not
possible. Instead, an indirect method was used to infer its value. The
assumption was made that the rate of the neutral events was
azimuthally symmetric ({\em i.e.}, that each main detector, including
MD7, experienced the same neutral-particle yield). The neutral
fraction in MD7 was then larger relative to the unblocked detectors
due to the suppression of the charged-particle yields (electrons and
pions) in the Pb wall. Again, using Geant4 simulations to determine
the suppression factors for electrons and pions due to the wall, the
resulting neutral fraction was found to be $f_{\rm NB}^7 = (51 \pm
9)$\%.  This relatively large neutral fraction in MD7, compared to
that in the unblocked detectors, arises mainly due to the factor of 70
suppression in the light yield for electrons caused by the Pb wall,
and the factor of 1.6 suppression in the smaller light yield from
pions. The neutral backgrounds do not arrive from the same direction
as the charged particles, and are mainly not intercepted by the
Pb wall, and so are largely unattenuated in MD7.
Further details on the neutral-background extraction can be found in
Ref.~\cite{Dissertation:Jim}.

\subsection{Asymmetry determination}\label{sec:asymmetry}

For Integrating-Mode data, the raw asymmetry as measured by a given
detector was calculated, for each helicity quartet, using:
\begin{equation}
  \label{eq:A_raw}
      A_{\rm raw} = \frac{Y_+ - Y_-}{Y_+ + Y_-} \; .
  \end{equation}
Here $Y_{\pm} = S_{\pm}/I_{\pm}$ is the detector yield, defined as
the integrated signal from a given detector $S_{\pm}$ (after subtraction
of the electronic pedestal) normalized to
the beam current $I_{\pm}$ in each helicity window.
In Eq.~\ref{eq:A_raw}, $Y_{\pm}$ is averaged over
the two positive (negative) helicity windows in a quartet.

The raw asymmetries $A_{\rm raw}$ were then corrected for several sources
of false asymmetry. These included (i) false asymmetries arising from
helicity-correlated fluctuations in the properties (trajectory and energy)
of the electron beam,  $A_{\rm beam}$, (ii) asymmetries arising
from interactions of the electron beam with a collimator in the beamline
(downstream of the target),
which we refer to as the beamline background, $A_{\rm BB}$, and (iii)
asymmetries caused by re-scattering in the pre-radiators upstream of
each detector, which we refer to as the re-scattering bias, $A_{\rm bias}$.
Each of these are discussed in turn below. \\

(i)  $A_{\rm beam}$: The helicity-correlated beam correction was determined via
\begin{equation}
  \label{eq:A_beam}
      A_{\rm beam} = - \sum\limits^5_{i=1} \left( \frac{\partial A}{\partial \chi_i} \right) \Delta \chi_i \; ,
  \end{equation}
where $\Delta \chi_i$ are the helicity-correlated differences in the
trajectory or energy, as measured over a helicity quartet.  The
sensitivities $\partial A / \partial \chi_i$ were determined during 6
minute intervals, using linear least-squares regression of the natural
fluctuations of five beam properties: position and angle in $x$ and $y$, and
energy. These corrections were applied to the data for each detector, for
each helicity quartet.  The net result of these corrections for each
detector was small, typically $< 0.05$~ppm, as detailed in
Tab.~\ref{tab:RegressedAsymmetries} and Tab.~\ref{tab:RegressedAsymmetriesTrans} , and the statistical uncertainty on these
corrections was negligible. \\

(ii) $A_{\rm BB}$: In the weak-charge measurement~\cite{Qweak.Nature},
it was found that a false asymmetry arose due to secondary events
scattered from the beamline and beam collimator. Such events were
determined to be predominantly comprised of low-energy neutral
particles which contributed a small amount to the detector signal, but
which carried a significant asymmetry, associated with
helicity-dependent intensity and/or position fluctuations in the halo
around the main accelerated electron-beam. This asymmetry was
monitored and corrected for, using the asymmetries measured in various auxillary
``background'' detectors (see Refs.~\cite{Qweak.Nature},
\cite{Dissertation:Manolis} for details).  The same technique was
adopted in the present analysis, resulting in only a small correction,
consistent with zero: $A_{\rm BB} = - 0.012\pm 0.027$~ppm~\cite{Dissertation:Jim}.\\

(iii) $A_{\rm bias}$: After the polarized electrons scattered from the
target, they traveled through the spectrometer's magnetic field,
causing their spins to precess. In the weak-charge
measurement~\cite{Qweak.Nature}, this resulted in the initially
longitudinally-polarized electrons developing a significant transverse
(radial) component upon reaching the main-detector array. These
electrons showered and underwent multiple scattering in the Pb
pre-radiators in front of the main detectors. The parity-conserving
left-right analyzing power in the low-energy Mott scattering of the
electrons from the Pb nuclei caused the asymmetries measured in the
two PMTs mounted on either end of a given main detector to differ. In
the weak-charge measurement, the difference between the asymmetries
for the two PMTs was found to be of the order of $A_{\rm diff} =0.3$~ppm.
Fortunately, for perfect detector symmetry, this
parity-conserving effect cancels when forming the parity-violating
asymmetry of interest. Small symmetry-breaking imperfections in the
main detector's geometry and optical response functions led to a
modest correction to the parity-violating asymmetry, which we refer to
as the re-scattering bias $A_{\rm bias}$. This effect was extensively
studied for the weak charge measurement~\cite{Qweak.Nature}, where the
correction was found to be $A_{\rm bias} = 0.0043 \pm
0.0030$~ppm. The re-scattering effect was also found in the present
data set; the difference of the asymmetries from the two PMTs,
averaged over all 8 detectors was $A_{\rm diff} = 1.3 \pm 0.3$~ppm.
The larger physics asymmetries, and larger statistical uncertainties
for the present measurement meant that a similarly detailed study of
$A_{\rm bias}$ was not required here. Instead, the previous value of
$A_{\rm bias}$ was simply scaled by the ratio of the $A_{\rm diff}$
values between the present measurement and that from the weak-charge
measurement, and, to be conservative, the uncertainty on $A_{\rm   bias}$
was doubled, to yield $A_{\rm bias} = 0.019 \pm 0.028$~ppm,
which is small compared to $A_{\rm raw}^{ii}$ (see Ref.~\cite{Dissertation:Jim} for details).

The raw asymmetries were corrected for the false asymmetries discussed above
in order to generate the measured asymmetries  $A_{\rm meas}^{ij}$ using
\begin{equation}
  \label{eq:A_meas}
      A_{\rm meas}^{ij} = A_{\rm raw}^{ij} + A_{\rm BB} + A_{\rm bias}  + A^{ij}_{\rm beam} \; .
  \end{equation}
Here, the index $i = 1,2...8$ represents the main-detector number, and
$j$ represents the data set, either mixed or transverse. The
same corrections $A_{\rm BB}$ and $A_{\rm bias}$ were used for all
8 detectors.  With eight main detectors and two data sets 
(mixed and transverse), there were sixteen total measured
asymmetries $A_{\rm meas}^{ij}$.  The raw and measured asymmetries for
the mixed data set are tabulated in
Tab.~\ref{tab:RegressedAsymmetries}, and for the transverse data set
in Tab.~\ref{tab:RegressedAsymmetriesTrans}.

A valuable test to ensure that false asymmetries have been properly accounted
for is the behavior of the asymmetries under the ``slow'' reversal, wherein an
insertable half-wave plate was periodically inserted into the path of the laser
beam in the electron-beam source. This reversed
the actual electron beam helicity with respect to the helicity signal from
the source, and so should simply switch the sign of the measured asymmetry.
A failure of this reversal would reveal the presence of several classes
of imperfectly corrected-for false asymmetries.

\begin{figure}[htb]
\begin{center}
\includegraphics[width=3.5in]{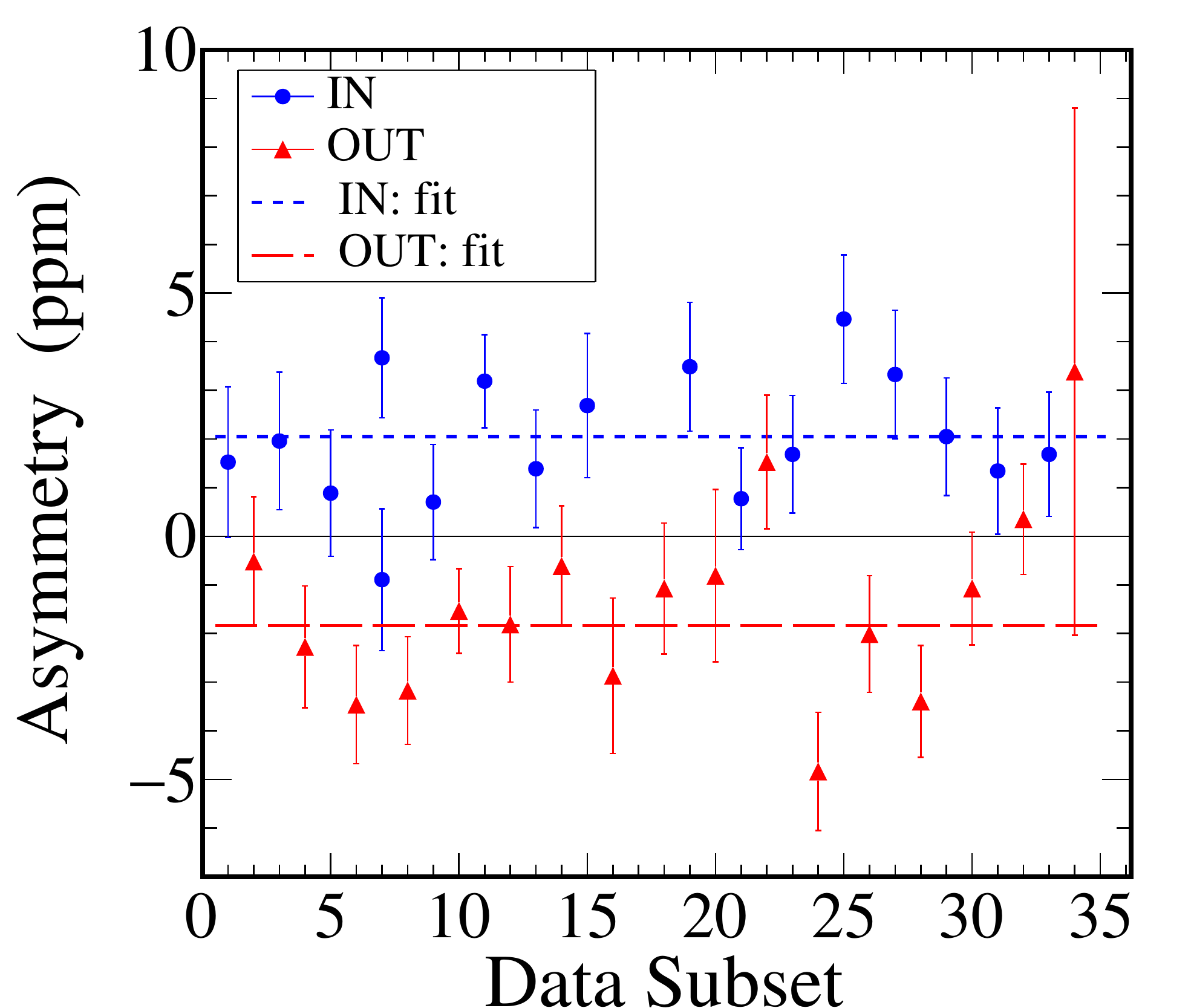}
\caption{Measured asymmetry from the mixed data-set, averaged over the unblocked detectors, {\em vs}.\ data subset,
 where each data subset corresponds to a particular
state of the insertable half-wave plate used to reverse the sign of the
electron-beam helicity. Each data subset represents roughly 4 hours of data-taking. The ``OUT'' subsets reveal the unreversed sign of the asymmetry.
} \label{fig:reversals}
\end{center}
\end{figure}

The measured asymmetries were well-behaved under the slow-reversal process.
For example, Fig.~\ref{fig:reversals} shows the average asymmetry from the 7 unblocked
detectors plotted {\em vs.}\ data subset, where each data subset corresponds to a particular
state of the insertable half-wave plate. The measured asymmetry reverses sign as expected; p-value for
a fit of the sign-corrected asymmetries to a single value is an acceptable
0.238.

\begin{table}[!hbt]
    \caption[Raw and Regressed Asymmetries]{Asymmetries
      for each main detector from the mixed data-set. Raw
      asymmetries, $A_{\rm raw}$, as well as the
      asymmetries after correction for helicity-correlated fluctuations in
      beam properties, $A_{\rm raw} + A_{\rm beam}$, are shown.
      Note that all corrections were less than 0.20~ppm and they
      caused no appreciable increase in uncertainty. Note too that MD7 was blocked with a lead wall to emphasize the pion signal. }
    \label{tab:RegressedAsymmetries}
    \centering
    \begin{tabular}{@{}ccc@{}}
        \toprule
        Main &  $A_{\rm raw}$ & $A_{\rm raw} + A_{\rm beam} $  \\
       Detector & (ppm) &  (ppm)  \\
      \hline 
        1 & $- 2.28 \pm 0.57$  & $- 2.24 \pm 0.57 $ \\
        2 & $- 2.24 \pm 0.57$  & $- 2.24 \pm 0.57 $ \\
        3 & $- 3.17 \pm 0.56$  & $- 3.19 \pm 0.56 $ \\
        4 & $- 2.54 \pm 0.58$  & $- 2.58 \pm 0.58 $ \\
        5 & $- 2.11 \pm  0.58$  & $- 2.10 \pm 0.58 $ \\
        6 & $0.35 \pm 0.58$  & $0.16 \pm 0.58 $ \\
        7 & $1.07 \pm  0.95$  & $1.07 \pm 0.95 $ \\
        8 & $- 1.46 \pm 0.57$  & $- 1.49 \pm 0.57 $ \\
        \hline
    \end{tabular}
  \end{table}

\begin{table}[!hbt]
    \caption[Raw and Regressed Asymmetries]{Asymmetries
      for each main detector from the transverse data-set. Raw
      asymmetries, $A_{\rm raw}$, as well as the
      asymmetries after correction for helicity-correlated fluctuations in
      beam properties, $A_{\rm raw} + A_{\rm beam}$, are shown.
      Note that all corrections were less than 0.15~ppm and they
      caused no appreciable increase in uncertainty. }
    \label{tab:RegressedAsymmetriesTrans}
    \centering
    \begin{tabular}{@{}ccc@{}}
        \toprule
        Main &  $A_{\rm raw}$ & $A_{\rm raw} + A_{\rm beam} $  \\
       Detector & (ppm) &  (ppm)  \\
      \hline 
        1 & $2.56 \pm 2.86$  & $2.58 \pm 2.87 $ \\
        2 & $6.10 \pm 2.85$  & $6.14 \pm 2.85 $ \\
        3 & $6.61 \pm 2.80$  & $6.58 \pm 2.80 $ \\
        4 & $2.77 \pm 2.88$  & $2.72 \pm 2.88 $ \\
        5 & $- 4.56 \pm  2.90$  & $- 4.50 \pm 2.90 $ \\
        6 & $- 1.07 \pm 2.88$  & $- 1.20 \pm 2.88 $ \\
        7 & $18.57 \pm  4.64$  & $18.61 \pm 4.64 $ \\
        8 & $- 3.87 \pm 2.85$  & $- 4.00 \pm 2.86 $ \\
        \hline
    \end{tabular}
  \end{table}

\subsection{Extraction of the Longitudinal Electron Asymmetry}\label{sec:pves}

The measured asymmetries $A_{\rm meas}^{ij}$ include contributions
from both scattered electrons and from pions generated in the target. For
each particle type, the asymmetry 
includes parity-violating contributions due to the longitudinal component
of the beam polarization, and parity-conserving contributions due to
the transverse component of the beam polarization. These latter asymmetries,
which are predominantly caused by two-photon exchange processes, vary
in a sinusoidal manner with the azimuthal location of the detectors.
Finally, the asymmetry measured in each detector was diluted by the fraction of the
yield arising from particles other than electrons and pions, which we
designate as a neutral background, as it was dominated by low-energy
gamma rays.

The sign convention for a parity-conserving asymmetry from transversely
polarized electrons is that the measured asymmetry varies as
$B_n \vec{P}\cdot \hat{n}$, where $B_n$ is the beam-normal single-spin
asymmetry, $\vec{P}$ is the electron spin-polarization vector, and
$\hat{n}=(\vec{k}\times\vec{k^{\prime}})/(|\vec{k}\times\vec{k^{\prime}}|)$
with $\vec{k} (\vec{k^{\prime}})$ being the momentum of the
incident (scattered) electron. In the present case, the electron
polarization is entirely in the horizontal plane, with positive
defined as beam-right. The scattered electron direction is encoded by
the azimuthal angle of the given main detector, $\phi^i$, with
$\phi =0^{\circ}$ defined as beam left, and $\phi$ increasing in a clockwise manner.
The azimuthal dependence of the asymmetry in this case reduces simply
to a function of $B_n \sin\phi^i$.

Thus, the measured asymmetries were parameterized as 
\begin{align}
  \begin{aligned}
    A_{\rm meas}^{ij}=P(1-f_{\rm NB}^i) \Big[(1-f_{\pi}^i)&(A_e^{L}\cos{\theta_P^j}+A_e^T\sin{\theta_P^j}\sin{\phi^i}) \\
    +f_{\pi}^i&(A_{\pi}^{L}\cos{\theta_P^j}+A_{\pi}^T\sin{\theta_P^j}\sin{\phi^i})\Big].
    \label{eq:Parameterized_Asymmetry}
\end{aligned}
  \end{align}
Here, $f_{\pi}^i$ is the fractional yield from pions; for MD7 this is $f_\pi^7$,
and for the seven unblocked main-detectors this is $f_\pi^{\rm avg}$.
$P = 0.870 \pm 0.006$ is the total polarization of the
    electron beam.  The longitudinal asymmetry from electrons (pions) is $A^L_{e(\pi)}$.  
  The transverse asymmetry from electrons (pions) is $A^T_{e(\pi)}$.  
  The beam polarization angle of data set $j$ is $\theta_P^j$, with
  $j$ = ``mix'' (mixed) or ``trans'' (transverse).   
  The neutral background yield-fraction for MD $i$ is $f_{\rm NB}^i$.  
  The fixed angles corresponding to the azimuthal angle placement of the main detectors are $\phi^i$, with $\phi^1 = 0^{\circ}, \phi^2 = 45^{\circ}$, etc.

\begin{figure}[thb]
\begin{center}
\includegraphics[width=3.7in]{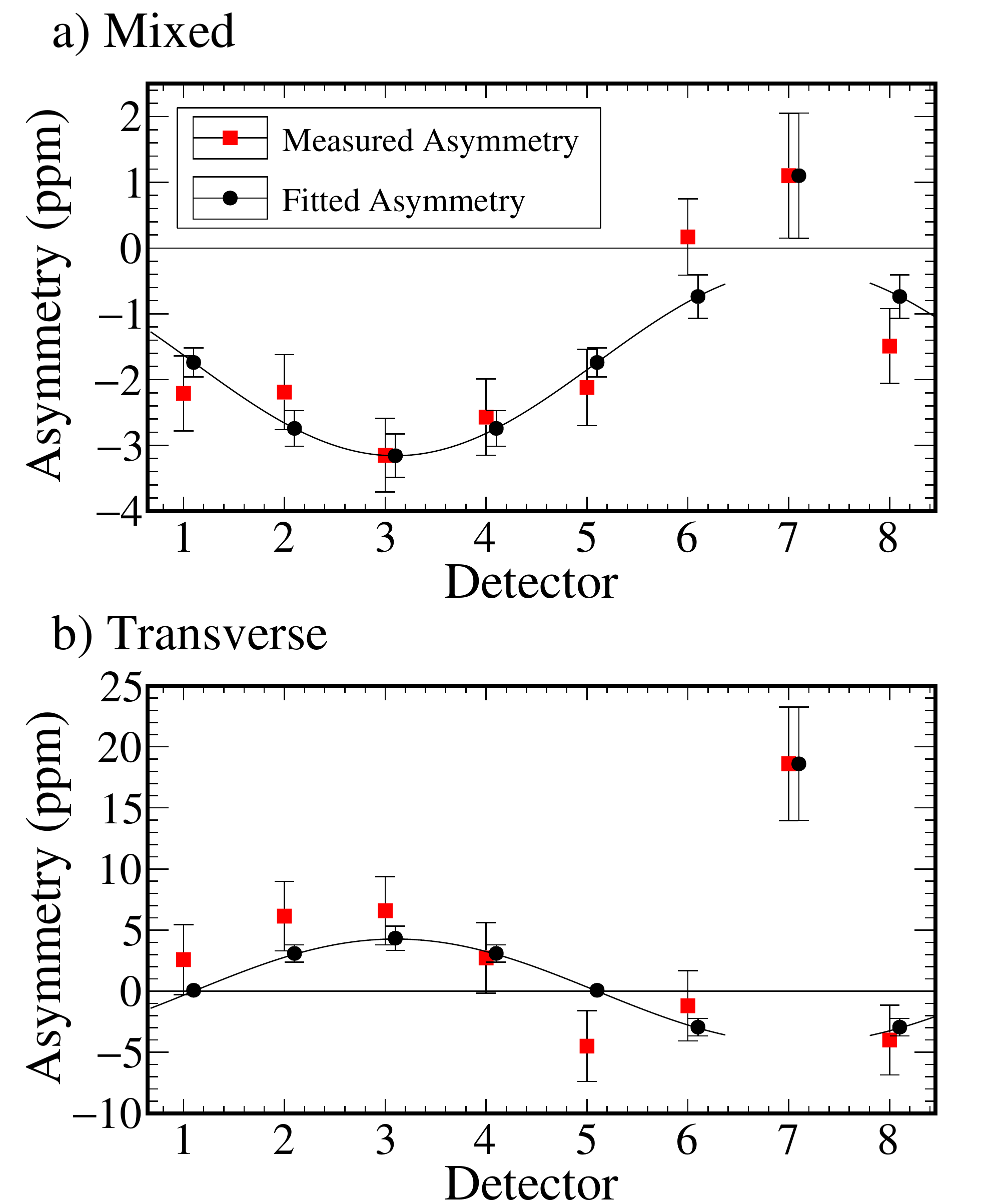}
\caption{Measured and fitted asymmetries {\em vs.}\ detector number,
  for both (a) the mixed data set and (b) the transverse data set. The fitted
  data points are connected via a line to guide the eye; Detector 7 is the
  ``blocked detector'' with enhanced sensitivity to pions.}
  \label{fig:asym_vs_octant}
\end{center}
\end{figure}

  In order to extract the four component asymmetries, $A^L_{e}$,
  $A^T_{e}$, $A^L_{\pi}$, and $A^T_{\pi}$, and their uncertainties from the measured
  asymmetries in Eq.~\ref{eq:Parameterized_Asymmetry}, a
  Monte-Carlo minimization approach was implemented.  The input
  quantities to this minimization were $A_{\rm meas}^{ij}$ (see Tab.~\ref{tab:RegressedAsymmetries}),
  $f_\pi^{\rm avg}$ (see Sec.~\ref{sec:pionyield}), $f_\pi^7$ (see Sec.~\ref{sec:pionmd7}), $f_{\rm NB}^i$ (see Sec.~\ref{sec:neutral}) and $\theta_P^j$. A value for each input quantity was randomly selected from a Gaussian
distribution about its mean with a standard deviation equal to its uncertainty. These random values were then used to calculate the asymmetry in each
MD and for each polarization configuration via 
\begin{align} 
\begin{aligned}
  A_{\rm calc}^{ij}=P(1-\tilde{f}_{NB}^i)\Big[(1-\tilde{f}_{\pi}^i)&(A_e^{L}\cos{\tilde{\theta}_P^j}+A_e^T\sin{\tilde{\theta}_P^j}\sin{\phi^i}) \\
      +\tilde{f}_{\pi}^i&(A_{\pi}^{L}\cos{\tilde{\theta}_P^j}+A_{\pi}^T\sin{\tilde{\theta}_P^j}\sin{\phi^i})\Big],
      \label{eq:A_calc}
\end{aligned}
\end{align}
where a `$\sim$' over a quantity indicates a randomly selected value for that quantity.
    The function $\delta$, where
    \begin{equation}
      \delta^2 = \sum_{i,j} (\tilde{A}_{\rm meas}^{ij}-A_{\rm calc}^{ij})^2 \; ,
    \end{equation}
 was then minimized with respect to the four unknown component asymmetries.
This resulted in one possible set of values for each component
asymmetry, $A^L_{e}$, $A^T_{e}$, $A^L_{\pi}$, and $A^T_{\pi}$.  The
randomization and minimization process was repeated $10^6$ times,
giving $10^6$ extracted values for each of the four component
asymmetries and $10^6$ values for the calculated asymmetries (shown in
Fig.~\ref{fig:asym_vs_octant}).  Iterating $10^6$
times ensured that each input quantity was sampled sufficiently to
span its full probability distribution.  This large amount of repeated
input sampling also ensured that the distributions of the extracted
component asymmetries varied smoothly.

The root-mean-squared (RMS) deviations of the resulting distributions
are taken as the uncertainties.  Correlations in the uncertainties
of the extracted quantities due to the fitting are automatically
accounted for in the Monte Carlo approach. The results for the four component
asymmetries and their uncertainties are listed in
Tab.~\ref{tab:Extracted_Asymmetries}.

    \begin{table}[htb]
      \caption[Table of extracted asymmetries]{Asymmetries extracted from the
        Monte-Carlo minimization process.}
      \label{tab:Extracted_Asymmetries}
      \centering
      \begin{tabular}{@{}ccc@{}}
          \toprule
          Asymmetry & ~~Value (ppm) \\
          \hline
          $A_e^L$ & -5.25 $\pm$ 1.49 \\
          $A_e^T$ & 12.3 $\pm$  3.6 \\
          $A_{\pi}^L$ & 25.4 $\pm$  9.0 \\
          $A_{\pi}^T$ & -60.1 $\pm$ 19.3 \\
          \hline
      \end{tabular}
    \end{table}

\subsection{Isolation of the Parity-Violating Inelastic Asymmetry}\label{sec:pvis}
  
The asymmetry of interest  $A_{\rm Inel}$  was contained
within the longitudinal electron asymmetry, $A_e^L$, which was determined as
described in the previous section. However $A_e^L$ needed to be
corrected for several physics backgrounds, and for the fact that the
beam was not 100\% polarized.  There were three significant such
background processes: (i) events generated by scattering in the
aluminum entrance and exit windows of the target, (ii) electrons
elastically scattered from the hydrogen that were radiated into the
acceptance and (iii) electrons elastically scattered from the hydrogen
that did not undergo radiation, but that generated signals in the detectors by
``punching through'' the concrete walls of the main-detector shielding-hut.

In order to correct for each of these backgrounds one needs to determine
the background fraction $f_k$, {\em i.e.}, the fraction of the signal
due to background $k$, as well as the asymmetry due to that background, $A_k$.
Below we discuss each of these backgrounds in turn. 

\subsubsection{Aluminum background}\label{sec:al}
\begin{figure}[htb]
\begin{center}
\includegraphics[width=3in]{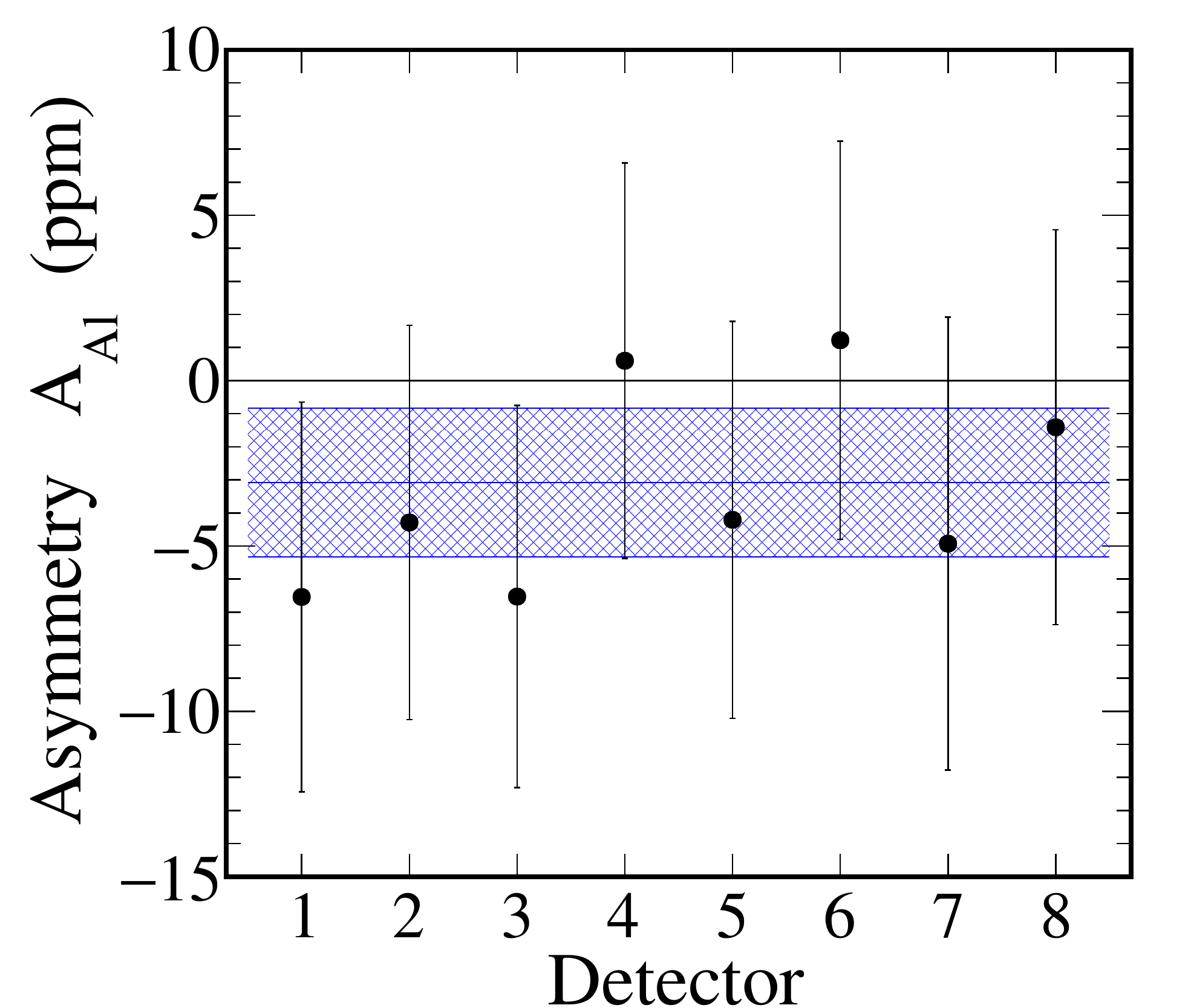}
\caption{$A_{\rm Al}$, the asymmetry measured from the aluminum alloy target,
 plotted {\em vs.}\  detector number. Errors shown are statistical only. The horizontal solid line and hatched region
  represent the average value and its uncertainty, respectively.}\label{fig:al_vs_octant}
\end{center}
\end{figure}

The fractional light-yield contribution from
the aluminum target-windows, $f_{\rm Al}$, was estimated using Geant4
simulation, yielding 
  \begin{equation}
    f_{\rm Al} = \frac{Y_{\rm Al}}{Y_{\rm Tot}} = 0.0075 \pm 0.0009 \;  , 
  \end{equation}
  where $Y_{\rm Al}$ is the yield of electrons scattered from the aluminum windows and $Y_{\rm Tot}$ is
  the total electron-yield from the cryogenic target. The cross-section parameterizations used
  in the simulation were taken from Refs. \cite{Christy:2007ve,PhysRevC.77.065206}. The longitudinal parity-violating asymmetry from aluminum $A_{\rm Al}^{\rm meas}$ was
  measured from dedicated runs taken on the aluminum-alloy target at the same
  beam energy. Consistent results for this asymmetry were measured 
  for all 8 detectors (see Fig.~\ref{fig:al_vs_octant}); the average value was $A_{\rm Al}^{\rm meas} =  -3.1 \pm 2.2$~ppm. Correcting for the beam polarization
 gives $A_{\rm Al} = A_{\rm Al}^{\rm meas}/(P\cos\theta_P^{\rm mix}) = -3.8 \pm 2.7$~ppm.

This asymmetry, for the unblocked detectors, contains contributions from
both electrons scattered from the aluminum as well as pions produced
in the aluminum. In principle, the pion and electron asymmetries from
aluminum could be separated by comparing the results for the unblocked
(electron-dominated) and blocked (pion-dominated) detectors. However,
there was not sufficient statistical precision in the aluminum-target data to
perform such a separation; the measured blocked detector (MD7)
asymmetry was identical within errors with that from the unblocked
detectors (see Fig.~\ref{fig:simple_adc_md7}). Consequently, we conservatively adopted the 8-detector
average with a doubled uncertainty, $A_{\rm Al} = - 3.8 \pm 5.4$~ppm as our
electron asymmetry from the Al windows. 
     
\subsubsection{Elastic Radiative Tail}\label{sec:tail}
Elastically-scattered electrons from the
target, if they did not undergo radiation, were too energetic to make it cleanly into the
acceptance of the spectrometer. However, if they underwent hard
bremsstrahlung, either before or after the scattering vertex, they
could emerge with energies ($\approx 1.0$ -- 1.2 GeV) that allowed them
into the acceptance.

The asymmetry of that fraction of the elastic radiative-tail that made
it into the acceptance and generated signals in the main detectors was determined using a Geant4
simulation. Elastically-scattered electrons were generated and propagated
through the target and the spectrometer; the asymmetries were
generated using the Standard Model value for the weak charge of the
proton (which is consistent with our measured value~\cite{Qweak.Nature}),
and the appropriate kinematics. External
bremsstrahlung processes were simulated using Geant4 routines, and
internal bremsstrahlung was accounted for following the prescription
of Schwinger~\cite{Schwinger:1981zz}. The acceptance-averaged
asymmetry extracted from the simulation was $A_{\rm El} = - 0.58 \pm
0.02$~ppm.

To calculate the background fraction $f_{\rm El}$ for this process, one
needs to know cross sections for highly-radiated elastic scattering,
but also those for the inelastic scattering-processes which represent
our signal of interest.

For radiative processes, following Mo and Tsai~\cite{Mo:1968cg} and
Tsai~\cite{Tsai:1973py}, the angle-peaking approximation was used when
calculating the angular integration of the cross sections, and the
equivalent-radiator approximation was used to calculate the
internal-bremsstrahlung corrections.  A Coulomb-correction was included
following Aste {\em et al.}~\cite{Article:Aste2005}.  For inelastic
scattering, the cross-section parameterization of Christy and
Bosted~\cite{Christy:2007ve} was adopted. The calculations of the
necessary radiative corrections to the cross sections were too
computationally expensive to directly embed in the Geant4
simulation. Instead, the cross sections were calculated using an
external piece of computer code originally developed by
S. Dasu~\cite{Thesis:Dasu1988}, and modified by a number of
authors. These calculated cross-sections were used to weight Geant4
simulated events, which thereby accounted for the experimental
acceptance. The resulting elastic radiative-tail yield-fraction was
$f_{\rm El} = Y_{\rm El}/Y_{\rm Tot} = 0.616 \pm 0.036$, where
$Y_{\rm El}$ is the yield of elastically scattered electrons.

\subsubsection{Elastic punch-through}\label{sec:pt}
The detector-hut shield-wall was designed for the primary weak-charge
measurement which had a maximum energy of scattered electrons $E' <
1.16$~GeV.  For the present measurement, with the beam energy tripled
to 3.3 GeV, the copious flux of elastically-scattered electrons was
dumped onto the shield wall. The $E'$ for elastic-scattering events
could reach near 3.3 GeV, and thus the concrete shield-wall was not
thick enough to absorb all of the energy for the most energetic of
these events. An additional background arose from those events which showered in
the shield wall, when some of the secondaries in the shower ``punched through''
and deposited light in the main detectors. 

\begin{figure}[htb]
\begin{center}
\includegraphics[width=3in]{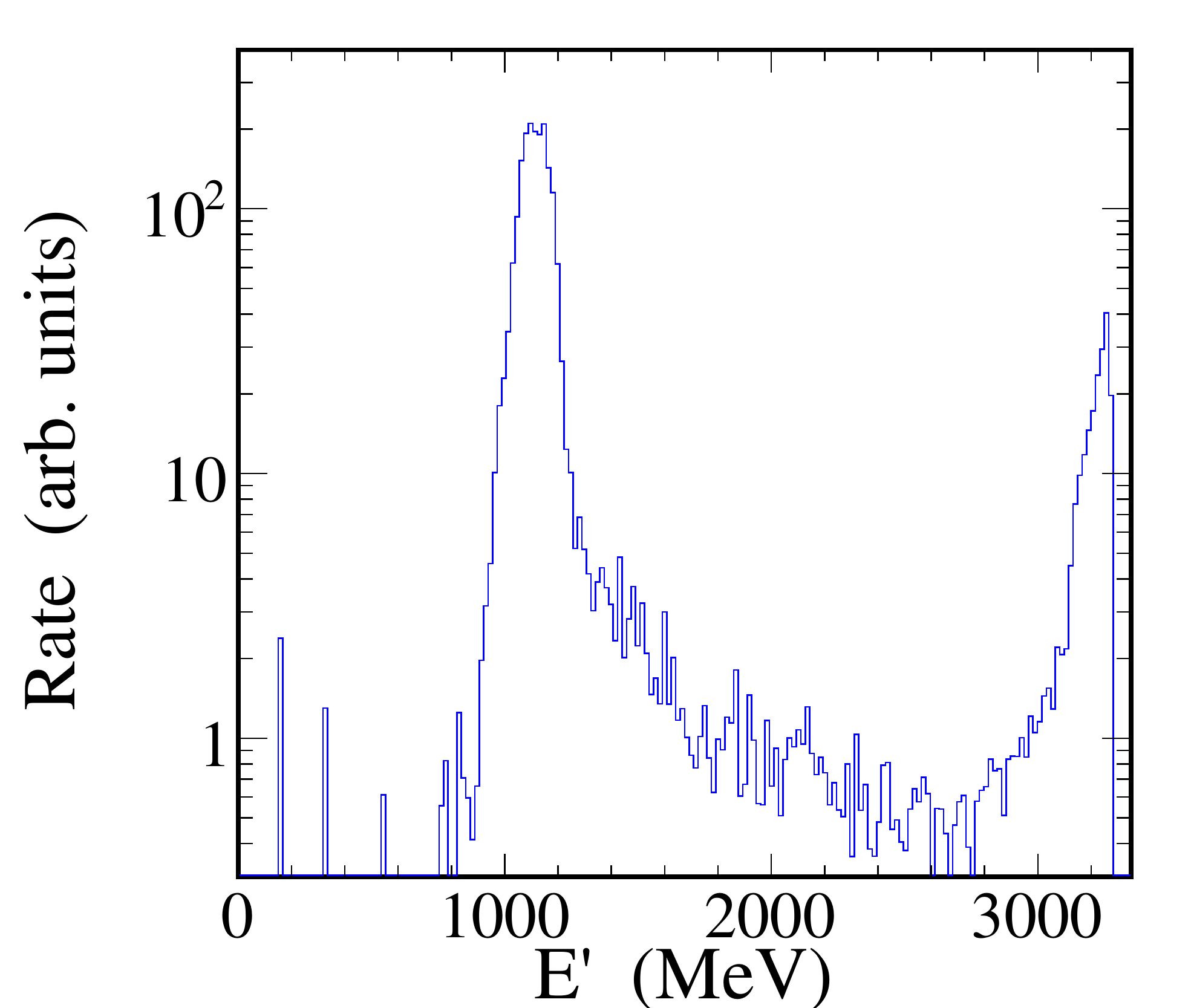}
\caption{Simulated energy $E'$ spectrum for elastically-scattered electrons
  for those events that generated signals in the main detectors (log
  scale). Note the two distinct peaks.  The peak near 1100 MeV
  represents highly-radiated scattered electrons that pass through the
  collimators and the apertures of the concrete shield-hut, to
  directly impinge on the main detector.  The peak at the right is due
  to elastically-scattered ``punch-through'' electrons (see text)
  which had radiated little energy, but which struck the shield-hut
  wall, creating a shower that generated signal in the main
  detectors.}\label{fig:eprime}
\end{center}
\end{figure}

To correct for this effect, Geant4 was used to simulate elastically
scattered electrons with scattered energies ranging from 150 MeV up
to 3.35 GeV.  Figure~\ref{fig:eprime} shows the energy spectrum of scattered
electrons for events for which light was deposited
in one of the the main detectors. The pulse-height weighted yield-fraction for these punch-through events obtained from the
simulation was $f_{\rm PT} = Y_{\rm PT}/Y_{\rm Tot} =  0.0220 \pm  0.0007$, where
$Y_{\rm PT}$ is the yield of these punch-through events, and their asymmetry
 was $A_{\rm PT} = -3.96\pm 0.04$~ppm.

\subsection{Central Kinematics}
 A direct measurement of the central value of the four-momentum transfer for the inelastic
events, $Q^2_{\rm Inel}$, using the tracking
system was not possible, because the inelastic events could not be
distinguished experimentally from the events in the elastic radiative-tail.
However, a value for $Q^2_{\rm total}$, the $Q^2$ for the
predicted mixture of inelastic and radiative-tail events, was extracted
from the Geant4 simulation. This simulated value, $Q^2_{\rm total (sim)} = 0.0787 \; {\rm GeV}^2$ was in reasonable agreement with the
experimental value from the tracking system,
$Q^2_{\rm total (meas)} = 0.0762 \; {\rm GeV}^2$, and the discrepancy between the
two was used to estimate the uncertainties on  $Q^2_{\rm Inel}$ and $W$.
The resulting four-momentum transfer was $Q^2_{\rm Inel} = 0.082 \pm 0.002 \; {\rm GeV}^2$, and the invariant mass $W = 2.23 \pm 0.06$ GeV.

\section{Results}\label{results}

   With all backgrounds having been measured or simulated, the final
parity-violating asymmetry from inelastic electron-proton
scattering $A_{\rm Inel}$ was extracted from $A_e^L$ using
    \begin{equation}
    \label{eq:Physics_Asymmetry}
        A_{\rm Inel} = \frac{A_e^L-\sum\limits_k f_k A_k}{1-\sum\limits_k f_k} \; ,
    \end{equation}
    where  $k = [{\rm El, PT, Al}]$,  corresponding to the
    elastic radiative-tail, elastic punch-through, and aluminum
    target-window backgrounds, respectively.

This physics asymmetry was determined to be
\begin{align}
\begin{aligned}
  A_{\rm Inel} &= -13.5\pm2.0 ({\rm stat.})\pm3.9 ({\rm syst.})\; {\rm ppm}  \\
                 &= -13.5\pm4.4 \; {\rm ppm}
                 \label{eq:A_phys_Result}
\end{aligned}
\end{align}
at $Q^2 = {0.082}~{\rm GeV}^2$ and $W = 2.23~{\rm GeV}$.
    
 The uncertainty of the final $A_{\rm Inel}$ was dominated by
 systematic uncertainties ($28.7\%$ relative) (see
 Tab.~\ref{tab:Uncertainty_Contributions}).  The four primary contributors, in decreasing order of size, were the
 pion yield-fractions $f_{\pi}$, the neutral background in MD7 $f_{\rm NB}^7$,
 the elastic radiative-tail yield-fraction $f_{\rm El}$, and the polarization angles of the
 electron beam $\theta_P$.

    \begin{table}[htb]
      \caption[]{Summary of contributions to $A_{\rm Inel}$, and their contributions to the uncertainty  $A_{\rm Inel}$, in relative percent.}
      \label{tab:Uncertainty_Contributions}
      \centering
      \begin{tabular}{@{}llcc@{}}
          \toprule
                                                &                   &                            & Contribution\\
          Quantity                              & Quantity             & Value                      & to $(\frac{dA_{\rm Inel}}{A_{\rm Inel}})$ \\
          \hline
          Pion Yield-Fraction                   & $f_\pi^{\rm avg}$     & $0.096 \pm 0.029$          & $21.9\%$ \\
                                                & $f_\pi^7$         & $0.81 \pm 0.05$            & \\
          Neutral Background in MD7             & $f_{\rm NB}^7$        & $0.51 \pm 0.09$            & $12.4\%$ \\
          Elastic Radiative-Tail Yield-Fraction & $f_{\rm El}$          & $0.62 \pm 0.04$            & $9.8\%$ \\
          Polarization Angle                    & $\theta_P^{\rm mix}$ & $-19.7^{\circ} \pm 1.9^{\circ}$ & $9.3\%$ \\
                                                & $\theta_P^{\rm trans}$& $92.2^{\circ} \pm 1.9^{\circ}$ & \\
          Neutral Background in Unblocked MDs   & $f_{\rm NB}^{\rm Un}$     & $0.063 \pm 0.006$          & $1.4\%$ \\
          Aluminum Window Asymmetry             & $A_{\rm Al}$          & $-3.8 \pm 5.4 ~{\rm ppm}$   & $1.0\%$ \\
          Beam Polarization                     & $P$               & $0.870 \pm 0.006 $          & $0.8\%$ \\
          Elastic Radiative Tail Asymmetry      & $A_{\rm El}$          & $- 0.58 \pm 0.02 ~{\rm ppm}$  & $0.3\%$ \\
          Re-scattering Bias Effect              & $A_{\rm bias}$        & $0.019 \pm 0.028 ~{\rm ppm}$      & $0.2\%$ \\
          Aluminum Window Yield-Fraction        & $f_{\rm Al}$          & $0.0075 \pm 0.0009$        & $0.2\%$ \\
          Punch-Through Yield-Fraction          & $f_{\rm PT}$          & $0.0220 \pm 0.0007$        & $<0.1\%$ \\
          Beamline Background Asymmetry         & $A_{\rm BB}$          & $0.012 \pm 0.027 ~{\rm ppm}$ & $<0.1\%$ \\
          Punch-Through Asymmetry               & $A_{\rm PT}$          & $-3.96 \pm 0.04 ~{\rm ppm}$ & $<0.1\%$ \\
          Regression Correction                 &                   & $<0.20 \pm 0.00 ~{\rm ppm}$ & $<0.1\%$ \\

          \hline
          Total Systematics & & & $28.7\%$ \\
          Statistics & & & $15.8\%$ \\
\hline
          \textbf{Total:} & & & \textbf{$32.8\%$} \\
      \end{tabular}
    \end{table}

\begin{figure}[htb]
\begin{center}
\includegraphics[width=3in]{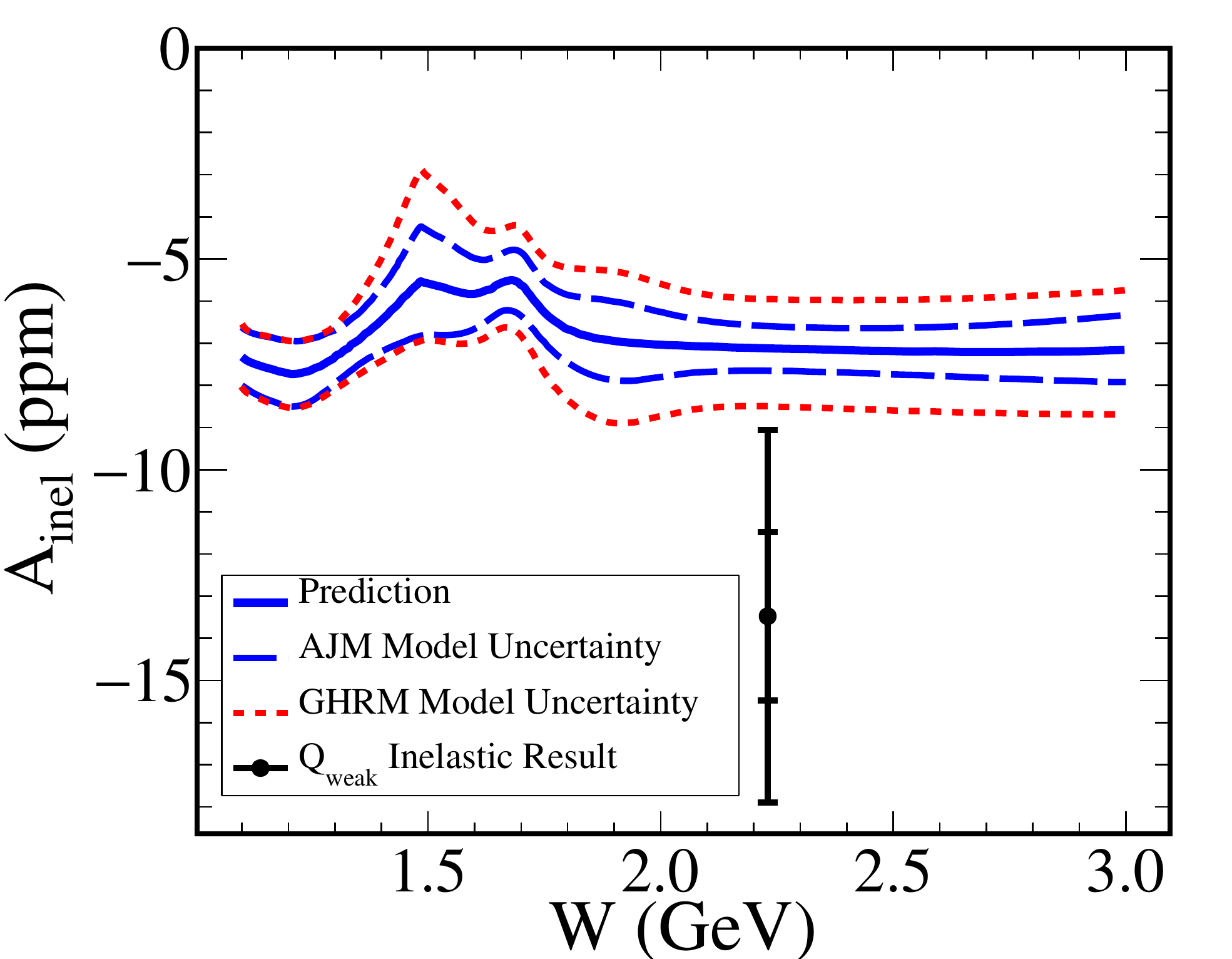}
\caption{Model calculations for $A_{\rm Inel}$ vs.\ $W$, at $Q^2 =  0.082$~GeV$^2$, and the present measured datum (solid circle). The
  central values (solid curve) are from the AJM group
  \cite{Hall:2013hta}, and the dashed lines represent the theoretical
  uncertainty from that calculation. The dotted lines are the
  larger theoretical uncertainties using the approach of Gorchtein {\em et
    al.} model \cite{Gorchtein:2011mz} (adapted from Fig.~16 of
  Ref. \cite{Hall:2013hta}). The statistical (inner) and total (outer)
  error bars for the present measurement are indicated. 
 } \label{fig:theory}
\end{center}
\end{figure}

The present result for $A_{\rm Inel}$ is compared to the predictions from
the AJM group \cite{Hall:2013hta} and from Gorchtein {\em et al.}
\cite{Gorchtein:2011mz} in Fig.~\ref{fig:theory}. Our central value is
larger in magnitude than that predicted by either of the two calculations, however
it agrees with both within $1.4 \sigma$ (experimental uncertainty).
The relatively limited precision of the present result does not allow us to
comment on the appropriateness of the somewhat smaller theoretical uncertainty
quoted by the AJM group (compared to that obtained by Gorchtein {\em et al.}).
Nonetheless, the agreement with both calculations lends confidence
in the modeling of the $F_1^{\gamma Z}$ and $F_2^{\gamma Z}$ interference structure functions used in these calculations, which are so critical to the $\Box_{\gamma Z}$ contribution to parity-violating electron scattering.

\subsection{Implications of other measured asymmetries}\label{sec:other}

In addition to the inelastic parity-violating asymmetry for electrons,
$A_{\rm Inel}$, which was the primary motivation for this measurement,
three other asymmetries were obtained from fitting the data (see
Tab.~\ref{tab:Extracted_Asymmetries}). These were the parity-violating
asymmetry for produced pions, $A_{\pi}^L$, and the transverse, or
beam-normal single-spin asymmetries (BNSSA) for scattered electrons and
produced pions, $A_e^T$ and $A_{\pi}^T$, respectively.

While we did not have sufficient data  available to fully correct
for the physics backgrounds (elastic radiative-tail and aluminum
target-windows) for these observables, we can nevertheless comment below on some
implications of these asymmetries.

\subsubsection{Electron Transverse Asymmetry $A_e^T$}

The beam-normal single-spin asymmetry $A_e^T$ is found to be significant and
positive ($12.3 \pm 3.6$ ppm), which is opposite in sign to all such
BNSSA measured to date in {\em elastic} electron scattering
experiments
\cite{Wells:2000rx,Maas:2004pd,Armstrong:2007vm,Androic:2011rh,Abrahamyan:2012cg},
from the proton and from complex nuclei. The aluminum window
contribution to the measured $A_e^T$ must be small, as there is no
significant transverse asymmetry seen in our Aluminum-alloy target
data (see
Fig.~{\ref{fig:al_vs_octant}).
In order to correct for transverse
asymmetries in the elastic radiative-tail, one would need either BNSSA
data at the appropriate kinematics, or a reliable theoretical model
for the BNSSA in elastic scattering from the proton. The relevant
kinematics are (i) that of elastic scattering at 3.3 GeV and
$Q^2 \approx 0.21~{\rm GeV}^2$ (for both punch-through events, Sec.~\ref{sec:pt}, and events that radiated after the scattering vertex),
and (ii) that of elastic scattering at $\approx 1.1$~GeV and $Q^2 \approx
0.05$~GeV$^2$ (for events that radiated before the scattering
vertex). Fortunately, there are data available for the elastic BNSSA
from the proton near both these kinematics.  Using a 3.0 GeV beam, the
G0 forward-angle transverse measurement~\cite{Armstrong:2007vm}
obtained $A_e^T = -4.1 \pm 1.2$ ppm at $Q^2 = 0.15$ GeV$^2$, and
$A_e^T = -4.8 \pm 2.1$ ppm at $Q^2 = 0.25$~GeV$^2$.  Our own
collaboration has a preliminary elastic result of $A_e^T = -5.5$~ppm at
1.16 GeV beam energy and Q$^2 = 0.025$~GeV$^2$ (Ref.~\cite{Waidyawansa:2013yva},
and to be published). Correcting the measured
$A_e^T$ for these asymmetries, weighted by their relative
contributions in a manner similar to that outlined in Sec.~\ref{sec:tail},
leads to a crude estimate of the purely-inelastic $A_e^T$ of $\approx +22$~ppm.

To our knowledge, there is no calculation available to date for the
BNSSA for the present inelastic kinematics ($Q^2 = {0.082}~{\rm
  GeV}^2$ and $W = 2.23~{\rm GeV}$),  which is above
the resonance region. However, Carlson {\em et al.}
\cite{Carlson:2017lys} have investigated the asymmetry for inelastic
scattering to the $\Delta(1232)$ resonance, at 1.16 GeV beam energy
and forward scattering-angle, and predicted large (40 -- 50~ppm)
positive values. That prediction is in good agreement with a
preliminary result at this beam energy from our collaboration
(Ref.~\cite{Nuruzzaman:2015vba}, and to be published). The Carlson {\em et
 al.} model included $\Delta(1232)$, $S_{11}(1535)$ and
$D_{13}$(1520) intermediate states in the hadron current. We speculate
that the large and positive asymmetry we have observed for inelastic
scattering above the resonance region may be driven by a similar mechanism
to that explored in the model of Carlson {\em et al.} 

\subsubsection{Pion Transverse Asymmetry $A_{\pi}^T$}

The measured transverse asymmetry $A_{\pi}^T = -60.1 \pm 19.3$ ppm can
provide information about the BNSSA in the inclusive production of
$\pi^-$'s. We did not attempt to correct the measurement for the
contribution from pions produced in the aluminum. However, under the
assumption that this contribution was small, and thus the signal is
dominated by pions produced from the hydrogen, then charge
conservation dictates that we are observing multiple mesons in the
final state. We are unaware of calculations appropriate to this
observable. We note, however, that Buncher and Carlson have calculated
the BNSSA in electron scattering in the resonance region when single
final-state hadrons are observed ~\cite{Buncher:2016nmv}. They point
out that in the case of inelastic processes in which only an outgoing
pion is observed, a BNSSA can arise from single-photon exchange
processes, via final-state strong interactions.  The generated BNSSA
can be of either sign, depending on kinematics \cite{Buncher:2016nmv}.

We note that our observation of such a large ($\approx -60$~ppm) BNSSA for
inclusive pions at multi-GeV beam energies is important for the design
of future precision parity-violation experiments such as the planned 11
GeV MOLLER experiment at Jefferson Lab \cite{Benesch:2014bas}, as
pions might produce significant azimuthally-varying
background asymmetries.

\subsubsection{Longitudinal Pion Asymmetry $A_{\pi}^L$}

A large positive asymmetry for parity-violating inclusive
$\pi^-$ production,  $A_{\pi}^L = 25.4 \pm 9.0$~ppm, was extracted
from the data. Again, we did not attempt to correct this for pions
produced in the aluminum windows so as to extract an asymmetry
for production from the proton. We can nevertheless make
some comments on this result. The parity-violating asymmetry in real
photoproduction should be of the order of the hadronic parity-violation
parameter $h^1_{\pi}$, which is experimentally known to be 
$(2.6 \pm 1.4) \times 10^{-7}$
\cite{Blyth:2018aon}. The asymmetry from electroproduction of
the $\Delta(1232)$ from protons or neutrons is negative, as measured
by G0 ~\cite{Androic:2012doa}. At much higher beam-energy
(50 GeV), the E158 collaboration observed a negative inclusive pion-production
asymmetry of order $-1$~ppm \cite{Anthony:2005pm}.

Explanation of the observed large {\em positive} asymmetry would seem to
require alternative physics mechanisms. One possibility is the
photoproduction (and electroproduction) of polarized hyperons, $\Lambda$ and $\Sigma$.
Large ($\approx 50\% - 75$\%) transfer of polarization has been observed in
electroproduction of hyperons at similar kinematics \cite{PhysRevC.79.065205}. The pion-emitting weak decay (eg $\Lambda \rightarrow p \pi^-$) of such hyperons is self-analyzing, and pions
emitted forward or backward in the hyperon rest-frame may have different
kinematic acceptances, thus leading to an asymmetry in the corresponding
detection efficiency.
A similar effect was seen in forward-angle parity-violating electron scattering
at 3.0 GeV beam energy by the G0 collaboration \cite{Armstrong:2005hs} who
found large, positive asymmetries for protons from hyperon decay.

An alternative hypothesis is that there are large contributions from
isoscalar exchange, {\em i.e.}, the virtual $Z^0$ producing multi-pion
final states and coupling to the nucleon via isoscalar Reggeon-exchange 
\cite{misha}. Another possible source would be pions produced in DIS at
large $W$. There is insufficient information in the present data to
distinguish between these possibilities. 

Again, our observation of large ($\approx +20$~ppm) inclusive
parity-violating pion asymmetries might be relevant as a source of
potential backgrounds for the MOLLER experiment~\cite{Benesch:2014bas}.

\section{Summary}\label{sec:summary}

We have measured the parity-violating asymmetry in the inelastic scattering
of electrons from the proton above the resonance region, at $Q^2 = {0.082}~{\rm GeV}^2$ and $W = 2.23~{\rm GeV}$. The result, $A_{\rm Inel} = -13.5\pm2.0 ({\rm stat.})\pm3.9 ({\rm syst.})\; {\rm ppm}$, is in acceptable  agreement
with two dispersion-model calculations \cite{Hall:2013hta} and \cite{Gorchtein:2011mz}. The measurement probes the low-$Q^2$ and high-$x$ kinematic region
that is most important for calculations of the $\Box_{\gamma Z}$
contribution to precision parity-violating electron-scattering
measurements. The result therefore lends confidence to these calculations
and to low-energy tests of the Standard Model that use them.
However, more precise measurements would be required in order to reduce the
uncertainty of the $\Box_{\gamma Z}$ calculations. 

We also observed a large positive BNSSA in inelastic electron-scattering,
a large negative BNSSA in the inclusive production of pions, and a
large positive asymmetry in the parity-violating inclusive production of
pions.

\begin{acknowledgments}
This work was supported by DOE Contract No. DEAC05-06OR23177, under
which Jefferson Science Associates, LLC operates Thomas Jefferson
National Accelerator Facility. Construction and operating funding for
the experiment was provided through the U.S. Department of Energy
(DOE), the Natural Sciences and Engineering Research Council of Canada
(NSERC), the Canadian Foundation for Innovation (CFI), and the
National Science Foundation (NSF) with university matching
contributions from William \& Mary, Virginia Tech, George Washington
University, and Louisiana Tech University. We thank the staff of
Jefferson Lab, in particular the accelerator operations staff, the
target and cryogenic groups, the radiation control staff, as well as
the Hall C technical staff for their help and support. We are grateful
for the contributions of our undergraduate students. We thank TRIUMF
for its contributions to the development of the spectrometer and
integrating electronics, and BATES for its contributions to the
spectrometer and Compton polarimeter. We are indebted to P.G. Blunden,
C.E. Carlson, M. Gorchtein, N.L. Hall, W. Melnitchouk, M.J. Ramsey-Musolf,
and A.W. Thomas for many useful discussions.
\end{acknowledgments}

\bibliographystyle{apsrev}
\bibliography{main.bib}

\end{document}